\documentclass[11pt,a4paper]{article}
\def\bea{\begin{eqnarray}}
\def\eea{\end{eqnarray}}
\def\nn{\nonumber}
\def\beq{\begin{equation}}
\def\eeq{\end{equation}}
\def\ba{\beq\new\begin{array}{c}}
\def\ea{\end{array}\eeq}
\def\be{\ba}
\def\ee{\ea}

\makeatletter
\newdimen\normalarrayskip              
\newdimen\minarrayskip                 
\normalarrayskip\baselineskip
\minarrayskip\jot
\newif\ifold             \oldtrue            \def\new{\oldfalse}
\def\arraymode{\ifold\relax\else\displaystyle\fi} 
\def\eqnumphantom{\phantom{(\theequation)}}     
\def\@arrayskip{\ifold\baselineskip\z@\lineskip\z@
     \else
     \baselineskip\minarrayskip\lineskip2\minarrayskip\fi}
\def\@arrayclassz{\ifcase \@lastchclass \@acolampacol \or
\@ampacol \or \or \or \@addamp \or
   \@acolampacol \or \@firstampfalse \@acol \fi
\edef\@preamble{\@preamble
  \ifcase \@chnum
     \hfil$\relax\arraymode\@sharp$\hfil
     \or $\relax\arraymode\@sharp$\hfil
     \or \hfil$\relax\arraymode\@sharp$\fi}}
\def\@array[#1]#2{\setbox\@arstrutbox=\hbox{\vrule
     height\arraystretch \ht\strutbox
     depth\arraystretch \dp\strutbox
     width\z@}\@mkpream{#2}\edef\@preamble{\halign
\noexpand\@halignto
\bgroup \tabskip\z@ \@arstrut \@preamble \tabskip\z@ \cr}%
\let\@startpbox\@@startpbox \let\@endpbox\@@endpbox
  \if #1t\vtop \else \if#1b\vbox \else \vcenter \fi\fi
  \bgroup \let\par\relax
  \let\@sharp##\let\protect\relax
  \@arrayskip\@preamble}
\def\eqnarray{\stepcounter{equation}%
              \let\@currentlabel=\theequation
              \global\@eqnswtrue
              \global\@eqcnt\z@
              \tabskip\@centering
              \let\\=\@eqncr
              $$%
 \halign to \displaywidth\bgroup
    \eqnumphantom\@eqnsel\hskip\@centering
    $\displaystyle \tabskip\z@ {##}$%
    \global\@eqcnt\@ne \hskip 2\arraycolsep
         $\displaystyle\arraymode{##}$\hfil
    \global\@eqcnt\tw@ \hskip 2\arraycolsep
         $\displaystyle\tabskip\z@{##}$\hfil
         \tabskip\@centering
    &{##}\tabskip\z@\cr}
\begingroup\ifx\undefined\newsymbol \else\def\input#1 {\endgroup}\fi
\input amssym.def \relax
\input amssym
\newfont{\hr}{msbm10}
\newfont{\ams}{msam10}

\font\numbers=cmss12
\font\upright=cmu10 scaled\magstep1
\def\stroke{\vrule height8pt width0.4pt depth-0.1pt}
\def\topfleck{\vrule height8pt width0.5pt depth-5.9pt}
\def\botfleck{\vrule height2pt width0.5pt depth0.1pt}
\def\Zmath{\vcenter{\hbox{\numbers\rlap{\rlap{Z}\kern 0.8pt\topfleck}\kern 2.2pt
                   \rlap Z\kern 6pt\botfleck\kern 1pt}}}
\def\Qmath{\vcenter{\hbox{\upright\rlap{\rlap{Q}\kern
                   3.8pt\stroke}\phantom{Q}}}}
\def\Nmath{\vcenter{\hbox{\upright\rlap{I}\kern 1.7pt N}}}
\def\Cmath{\vcenter{\hbox{\upright\rlap{\rlap{C}\kern
                   3.8pt\stroke}\phantom{C}}}}
\def\Rmath{\vcenter{\hbox{\upright\rlap{I}\kern 1.7pt R}}}
\def\Z{\ifmmode\Zmath\else$\Zmath$\fi}
\def\Q{\ifmmode\Qmath\else$\Qmath$\fi}
\def\N{\ifmmode\Nmath\else$\Nmath$\fi}
\def\C{\ifmmode\Cmath\else$\Cmath$\fi}
\def\R{\ifmmode\Rmath\else$\Rmath$\fi}

\newcounter{app}

\def\app{\setcounter{equation}{0}
\def\theequation{\Alph{app}.\arabic{equation}}\par
   \addvspace{4ex}
   \@afterindentfalse
  \secdef\@app\@dapp}

\newcommand\@app{\@startsection {app}{1}{0ex}%
                                   {-3.5ex \@plus -1ex \@minus -.2ex}%
                                   {2.3ex \@plus.2ex}%
                                   {\normalfont\Large\bf}}
\def\@dapp#1{%
{\parindent \z@ \raggedright  \bf #1}\par\nobreak}
\def\l@app#1#2{\ifnum \c@tocdepth >\z@
    \addpenalty\@secpenalty
    \addvspace{1.0em \@plus\p@}%
    \setlength\@tempdima{8em}%
    \begingroup
      \parindent \z@ \rightskip \@pnumwidth
      \parfillskip -\@pnumwidth
      \leavevmode \bfseries
      \advance\leftskip\@tempdima
      \hskip -\leftskip
      #1\nobreak\hfil \nobreak\hb@xt@\@pnumwidth{\hss #2}\par
    \endgroup\fi}
\newcounter{sapp}[app]

\def\sapp{\def\theequation{\Alph{app}.\arabic{equation}}
\par
\@afterindentfalse
  \secdef\@sapp\@dsapp}
\newcommand{\@sapp}{\@startsection{sapp}{2}{\z@}%
                                     {-3.25ex\@plus -1ex \@minus 
-.2ex}%
                                     {1.5ex \@plus .2ex}%
                                     {\normalfont\large\bfseries}}

\def\@dsapp#1{%
{\parindent \z@ \raggedright  \bf #1
}\par\nobreak}
\newcommand{\l@sapp}{\@dottedtocline{2}{1.5em}{2.3em}}

\catcode`\@=11

\def\section{\@startsection{section}{1}{\z@}{3.5ex plus 1ex minus
   .2ex}{2.3ex plus .2ex}{\large\bf}}

\def\thesection{\Roman{section}.}

\def\appendix{\setcounter{section}{0}
        \def\thesection{Appendix }
       \def\theequation{\Alph{section}.\arabic{equation}}}

\@addtoreset{equation}{section}
\def\theequation{\arabic{section}.\arabic{equation}}

\def\2{{1\over 2}}
\def\N2{${\cal N}=2$}

\def\be{ \begin{eqnarray} }
\def\ee{ \end{eqnarray} }

\def\bea{\begin{eqnarray}}
\def\eea{\end{eqnarray}}
\def\nn{\nonumber}

\def\beq{\begin{equation}}
\def\eeq{\end{equation}}
\def\ba{\beq\new\begin{array}{c}}
\def\ea{\end{array}\eeq}
\def\be{\ba}
\def\ee{\ea}

\title{Calculating Gluino-Condensate Prepotential}

\author{
H.Itoyama$^{1}$ and A.Morozov$^{2}$ \
\\ \normalsize \em $^{1}$ 
Department of Mathematics and Physics, Osaka City University, Japan
\\
\normalsize \em $^{2}$
Institute of Theoretical and Experimental Physics, Moscow}

\date{December, 2002}

\begin{document}

\maketitle

\vspace{-7.7cm}

\begin{center}
\hfill hep-th/0212032 \\
\hfill OCU-PHYS 196\\
\hfill ITEP/TH-59/02
\end{center}

\vspace{6.5cm}

\begin{abstract}
We discuss the derivation of the CIV-DV prepotential for arbitrary
power $n+1$ of the original superpotential in the $N=1$
SUSY YM theory (for arbitrary number $n$ of cuts in the
solution of the planar matrix model in the Dijkgraaf-Vafa
interpretation).
The goal is to hunt for structures, not so much for
exact formulas, which are necessarily complicated,
before the right language is found to represent them.
Some entities, reminiscent of representation theory,
clearly emerge, but a lot of work remains to be done
to identify the relevant ones. 
As a practical application, we obtain a cubic
(first non-perturbative) contribution to the prepotential
for any $n$.
\end{abstract}

\vspace{2.0cm}


After the recent papers of R.Dijkgraaf and C.Vafa \cite{DV}
the calculation of gluino-condensate prepotential \cite{CIV}
attracted wide attention, and many results are already
obtained \cite{followup}-\cite{IM5} for various examples, 
starting both from the planar matrix models solutions \cite{MaMop}
and from the Seiberg-Witten (SW) theory \cite{SW,GKMMM,IM4}.
However, not enough attention is yet devoted to the general
structure of the {\it multi-cut} prepotentials (with any number
$n$ of cuts, i.e. for any power $n+1$ of the superpotential,
i.e. for any genus $n-1$ of the spectral curve).
 ( Please note that the prepotential with $n$ cuts is valid for any 
  breaking/splitting of  the gauge group $U(N)$ into $n$ pieces
  $N= \sum_{i}^{n} N_{i}$ and does not depend on $N_{i}$.)
At the same time, the precise understanding of  this structure is
of crucial importance for establishing relations between
the Dijkgraaf-Vafa (DV) theory and other branches of
mathematical physics and string theory, such as 
the theory of effective actions in the spirit of \cite{UFN2},
representation theory, finite-$N$ matrix models \cite{UFN3},
KP/Toda-like $\tau$-functions \cite{tauf} 
and their generalizations \cite{qutauf}, 
instanton calculus {\it a la} refs.\cite{Ne,Flu,gra}, 
WDVV equations \cite{MMM,IM5} etc.

\section{The plan of calculation}

We want to evaluate the set of integrals

\be
\frac{1}{g_{n+1}} S_i = 2\int_{\gamma_i-\rho_i}^{\gamma_i+\rho_i}
dS_{DV}, \nn \\
\frac{1}{g_{n+1}} \Pi_i = 2\int_{\gamma_i+\rho_i}^\Lambda
dS_{DV}
\label{ints}
\ee
($i=1,\ldots,n$), which are the periods of the meromorphic (with poles
at infinities) differential

\be
dS_{DV} = y(x) dx
\ee
on the hyperelliptic complex curve (Riemann surface) of genus $n-1$
with quadratic branching points at $\gamma_i\pm\rho_i$:

\be
y^2(x) = \prod_{i=1}^n(x-\gamma_i-\rho_i)(x-\gamma_i+\rho_i)
= \prod_{i=1}^n((x-\gamma_i)^2-\rho_i^2)
\label{curve1}
\ee
The periods $\Pi_i$ are to be further interpreted as the $S_i$-derivatives
of the prepotential,

\be
\Pi_i = \left.\frac{\partial{\cal F}}
{\partial S_i}\right|_{constant\ \alpha's}
\ee
evaluated at constant values of parameters $\alpha_i$,
which are defined to be the roots of a polynomial

\be
P_n(x) = \prod_{i=1}^n (x-\alpha_i)
\ee
of degree $n$, such that the curve (\ref{curve1}) is also

\be
y^2(x) = P_n^2(x) + f_{n-1}(x),
\label{curve2}
\ee
where the polynomial 

\be
f_{n-1}(x) = \sum_{i=1}^n \tilde S_i \prod_{j\neq i}^n(x-\alpha_j) =
P_n(x)\sum_{i=1}^n \frac{\tilde S_i}{x-\alpha_i}
\ee
is of degree $n-1$ only.
According to the general arguments of the SW theory \cite{SW,GKMMM,IM4},
such system of equations is consistent, i.e.

\be
\left.\frac{\partial\Pi_i}
{\partial S_j}\right|_{constant\ \alpha's} =
\left.\frac{\partial\Pi_j}
{\partial S_i}\right|_{constant\ \alpha's}
\label{perma}
\ee
because the variation 

\be
\delta dS_{DV} = \sum_{i=1}^n \delta \tilde S_i dv_i(x)
+ O(\delta \alpha)
\ee
for $\delta \alpha_i=0$ is a linear combination of (almost)
holomorphic differentials

\be
dv_i(x) = \frac{\prod_{j\neq i}^n(x-\alpha_j)}{y(x)}dx
\ee
on the curve (\ref{curve1}). In fact $dv_i(x)$ possess
simple poles at $x=\infty_\pm$, but this does not spoil the
symmetric property of the period matrix (\ref{perma}) in the limit 
$\Lambda \rightarrow \infty$ (deviations from (\ref{perma})
are of the order $O(\alpha/\Lambda)$).

This means that the equations can be integrated and provide 
a function, the CIV-DV prepotential ${\cal F}(S|\alpha)$,
which has the form of

\be
2\pi i{\cal F}(S|\alpha) = 4\pi ig_{n+1}\left(W_{n+1}(\Lambda)\sum_i S_i 
- \sum_i W_{n+1}(\alpha_i)S_i\right) -
(\sum_{i} S_i)^2\log\Lambda +
\nn \\ +
\frac{1}{2} \sum_{i=1}^n S_i^2\left(\log \frac{S_i}{4}\ - \frac{3}{2}\right) 
- \frac{1}{2}\sum_{i<j}^n(S_i^2 - 4S_iS_j + S_j^2)\log \alpha_{ij} + 
\sum_{k=1}^\infty \frac{1}{(i\pi g_{n+1})^k}{\cal F}_{k+2}(S|\alpha),
\label{prepoexp}
\ee
where ${\cal F}_{k+2}(S|\alpha)$ are polynomial of degree $k+2$ in $S$'s
with $\alpha$-dependent coefficients, $\alpha_{ij} = \alpha_i - \alpha_j$
and $W_{n+1}'(x) = P_n(x)$.

The problem of calculating ${\cal F}_{k+2}(S|\alpha)$
is technically involved.\footnote{An 
advantageous approach to it
is provided by the DV interpretation of the same quantity 
in terms the planar large-$N$ limit of matrix models, which
allows to apply functional-integral techniques, in particular,
the usual QFT perturbation theory (diagram expansion).
}
Its solution includes 4 steps:\footnote{Note some technical change as
compared to s.5.3 of ref.\cite{IM4}. Such minor modifications
provide substantial technical simplifications.}

1) Evaluation of integrals (\ref{ints});

2) Inversion of formulas for $S_i(\rho,\gamma)$, providing expressions 
for $\rho(S,\gamma)$, and substitution of those into the 
$\Lambda_i$-independent ($\Lambda_i=\Lambda-\gamma_i$) 
pieces of integrals $\Pi_i$ (the $\Lambda_i$-dependent contributions
are better left expressed through $\gamma$ and $\rho$ at this stage);

3) Expression of $\gamma$'s through $\alpha$'s 
(and $\rho$'s) by comparison of (\ref{curve1}) and (\ref{curve2});

4) Substitution of the resulting expressions into those for
$\Pi$'s, integrating over $S$'s and obtaining the desired 
expansion for the prepotential ${\cal F}(S|\alpha)$.

In what follows we present some results which 
allow us to reveal some of the structures; they emerge at every step
of our procedure. There is of course, substantial overlap
with the previous considerations in \cite{DV}-\cite{IM5}.

A practical outcome of our calculation is an expression for the
first non-perturbative term in (\ref{prepoexp}):\footnote{
To match the overall coefficient to the one, frequently met in
the literature, one should rescale $S$: our $S/4$ looks equal to
$S$ in ref.\cite{CIV}.
}

\be
{\cal F}_3(S|\alpha) = \sum_{i=1}^n u_i(\alpha)S_i^3
+ \sum_{i\neq j}^nu_{i;j}(\alpha) S_i^2S_j +
\sum_{i<j<k}^n u_{ijk}(\alpha)S_iS_jS_k, \nn \\
u_i(\alpha) = \frac{1}{6}
\left(
-\sum_{j\neq i}\frac{1}{\alpha_{ij}^2\Delta_j} +
\frac{1}{4\Delta_i}\sum_{\stackrel{j<k}{j,k\neq i}}
\frac{1}{\alpha_{ij}\alpha_{ik}}
\right), \nn \\
u_{i;j}(\alpha) = \frac{1}{4}
\left(
-\frac{3}{\alpha_{ij}^2\Delta_i}
+ \frac{2}{\alpha_{ij}^2\Delta_j}
- \frac{2}{\alpha_{ij}\Delta_i}\sum_{k\neq i,j}\frac{1}{\alpha_{ik}}
\right), \nn \\
u_{ijk}(\alpha) = \frac{1}{\alpha_{ij}\alpha_{ik}\Delta_i} +
 \frac{1}{\alpha_{ji}\alpha_{jk}\Delta_j} +
 \frac{1}{\alpha_{ki}\alpha_{kj}\Delta_k}, 
\nn \\
\Delta_i = W_{n+1}^{\prime\prime}(\alpha_i) = \prod_{j\neq i}^n\alpha_{ij}
\label{F3}
\ee

In what follows, we begin with the step 1): evaluation of the integrals
(\ref{ints}) in terms of $\gamma$ and $\rho$. First, in s.2 we analyze
the expressions for these integrals through symmetric functions of 
$\gamma_{ij}$ (to be called $r(\gamma)$, $e(\gamma)$ and 
$\varepsilon(\gamma)$), certain
elementary integrals (to be called $J_m$, not to be mixed with Bessel
functions!) and peculiar infinite-order differential operator (to be called
$\hat D(\rho,\gamma)$, this one actually {\it is}  a Bessel function
of $\rho^2\partial_\gamma^2$). 
The main result of s.2 is eq.(\ref{prepexpansion30})
and/or eq.(\ref{prepexpansion3}), which may look more complicated,
but is more useful for practical calculations. Eqs.(\ref{prepexpansion30})
and (\ref{prepexpansion3}) are valid even for
{\it indefinite} integrals, expressions for concrete integration limits,
explicit in (\ref{ints}) are obtained by substitution from (\ref{JA})
and (\ref{JB}).
Then, in s.3 we concentrate on the first terms of expansion in powers
of $\rho^2$, relevant for evaluation of ${\cal F}_3$: this means
that we need $\rho^0$, $\rho^2$ and $\rho^4$. Then only a few types
of symmetric functions contribute, and the situation can be drastically
simplified.
In the same s.3 we perform step 2), in application to the 
$\Lambda_i$-independent terms in $\Pi_i$, and this leads to new
drastic simplifications. The $\Lambda_i$-dependent contributions to
$\Pi_i$ also get together into a simple expression, $W_{n+1}(\Lambda)-
W_{n+1}(\gamma_i)$, but to see this one needs to rely upon the
"$\gamma\ vs\ \alpha$ sum rules" (\ref{sumrules2}) from Appendix A.
The main result of  s.3 consists of simply-structured formulas
(\ref{uniterms}), (\ref{f3terms}) for $\Pi_i$. 
Eq.(\ref{uniterms}) contains  explicit $\gamma$-dependence 
(instead of being entirely $\alpha$-dependent) in just two places:
as an argument in $W_{n+1}(\gamma_i)$ and through $\log\gamma_{ij}$.
Expression for $\gamma$ through $S$ and $\alpha$, required
at the next step 3) are derived in Appendix A, the main result of which is
in eq.(\ref{sigmaS}), although the sum rules (\ref{sumrules2})
can sometime be directly applied as well, for example, to handle
the $\Lambda_i$-dependent terms in $\Pi_i$.
In s.4 we finally combine the results of s.3 and Appendix A to obtain
the final formulas (\ref{prepoexp}) and (\ref{F3}).
Appendix B contains sample calculations for $n=2$ and $n=3$, which can
be used for illustrative purposes.

\section{Periods through cuts}

\subsection{Notation and basic integrals}

For the purpose of evaluation of $S_i$ and $\Pi_i$
with given $i$, we need only to expand $dS_{DV}$ in powers of $\rho_j$
with all $j\neq i$, keeping the $\rho_i$ dependence exact in all terms of the
expansion. This is needed because the integrals in question are not
analytic in $\rho_i$ and are not evaluated by expansion of the
{\it integrands} in $\rho_i$. 

Namely, we express $dS_{DV}(x)/dx = \sqrt{{\cal P}_{2n}(x)}$
through $\gamma$'s and $\rho$'s in the following way:

\be
\frac{dS_{DV}(x)}{dx} =
\sqrt{{\cal P}_{2n}(x)} =
\sqrt{(x-\gamma_i)^2 - \rho_i^2}
\prod_{j\neq i}^n \sqrt{(x-\gamma_j)^2 - \rho_j^2}
\ee
Since we are further going to integrate over $x$, we can shift
the integration variable $x \rightarrow x_i+\gamma_i$,
$\gamma_{ij} = \gamma_i-\gamma_j$ (and remember
later to appropriately adjust the integration domains):

\be
\int_i dS_{DV}(x) =
\int dx \sqrt{x^2 - \rho_i^2}
\prod_{j\neq i}^n \sqrt{(x+\gamma_{ij})^2 - \rho_j^2} 
 = \nn \\ =
\int\sqrt{x^2 - \rho_i^2}\prod_{j\neq i}^n
\left\{
\sum_{m_j=0}^\infty c_{m_j} \rho_j^{2m_j}(x+\gamma_{ij})^{1-2m_j}
\right\}dx
 = \nn \\ =
\sum_{\{m_1,\ldots,m_n\}}
\left(\prod_{j\neq i} c_{m_j}\rho_j^{2m_j}\right)
I^{[i]}_{m_1,\ldots,m_n}(\rho_i;\gamma_{i\cdot})
\label{prepexpansion}
\ee
Here the coefficients\footnote{
Somewhat amusingly, just the same $c_m$ 
(or what is the same, square root expansions)
appear in many places, which seem unrelated
(though all can be traced back to the fact that we deal with
a hyperelliptic curve with quadratic branching points),
thus signaling some hidden symmetry of the problem
and probable short-cuts, overlooked in the presentation below.
The first few coefficients $c_m$ are:

$$
c_0=1,\ c_1=-1/2,\ c_2=-1/8,\ c_3=-1/16,\ c_4=-5/128,\
c_5=-7/256,\ c_6=-21/1024, \ldots
$$
}
$c_m=\frac{\Gamma (m-1/2)}{m!\Gamma(-1/2)}$
and integrals

\be
I^{[i]}_{m_1,\ldots,m_n}(\rho_i;\gamma_{i\cdot}) =
\int \sqrt{x^2 - \rho_i^2} \prod_{j\neq i}^n (x+\gamma_{ij})^{1-2m_j}dx.
\label{integral}
\ee
(Note that there are just $n-1$ indices $m_1,\ldots,m_n$ -- there is
nothing standing at the $i$-th position. Also by $\gamma_{i\cdot}$,
with a dot placed at the second index,
we denote the set $\{\gamma_{ij}\}$ with all possible values of $j\in \epsilon$,
see below.)

Whenever some $m_j\geq 1$, one can use the recurrent formula,

\be
I^{[i]}_{\ldots m_j\ldots}(\rho_i;\gamma_{i\cdot}) =
\frac{1}{(2m_j-2)!}
\left(\frac{\partial}{\partial \gamma_{ij}}\right)^{2m_j-2}
I^{[i]}_{\ldots 1_j \ldots}(\rho_i;\gamma_{i\cdot})
\label{recrels}
\ee

Having this formula in mind, it is reasonable to introduce more notation.

First, let us split the set of $n-1$ numbers
$N^{[i]}_n = \left\{1,2,\ldots,\check i,\ldots, n\right\}$ 
($i$ is omitted) into two non-intersecting sets:
$\epsilon$, consisting of all $k$ with $m_k=0$ and 
$\bar\epsilon$, including all $k$ with $m_k \geq 1$,

\be
N^{[i]}_n = \epsilon \cup \bar\epsilon, \ \
\epsilon \cap \bar\epsilon = \emptyset, \nn \\
\epsilon \equiv \left\{ k \in N^{[i]}_n: \ \ m_k = 0\right\}, \ \
\bar\epsilon \equiv \left\{ k \in N^{[i]}_n: \ \ m_k > 0\right\}
\ee
The numbers of elements in $\epsilon$ and $\bar\epsilon$ will
be denoted by $d(\epsilon)$ and $d(\bar\epsilon)$.
Somewhat later we will also need

\be
\epsilon_{j_1\ldots j_l} \equiv \epsilon/\left\{j_1,\ldots, j_l\right\}
= \left\{ k \in N^{[i]}_n: \ \ m_k = 0,\ k\neq j_1,\ldots, j_l\right\}
\ee
(assuming that $j_1,\ldots, j_l \in \epsilon$ and are all different),
and similarly

\be
\bar\epsilon_{j_1\ldots j_l} \equiv 
\bar\epsilon/\left\{j_1,\ldots, j_l\right\}
= \left\{ k \in N^{[i]}_n: \ \ m_k > 0,\ k\neq j_1,\ldots, j_l\right\}
\ee

Second, we denote by $\sum_{\epsilon,\bar\epsilon}$ the sum over all the
splittings of $N^{[i]}_n$ into two non-intersecting subspaces. The sum
contains $2^{n-1}$ terms (the number of subsets in $N^{[i]}_n$).

Third, the structure of the formulas (\ref{prepexpansion}) and
(\ref{recrels}), together with the relation

\be
(2m-2)!c_1 = 2^{2m-2}(m-1)!m!c_m, \ \ c_1 = -1/2,
\ee
imply the following definition of a formal infinite-order
differential operator

\be
\hat D(\rho;\gamma) = \sum_{m\geq 1}
\frac{c_m\rho^{2m}}{(2m-2)!}
\left(\frac{\partial}{\partial(\gamma)}\right)^{2m-2}
= c_1\sum_{m=0}^\infty
\frac{\rho^{2m+2}}{m!(m+1)!}
\left(\frac{\partial}{\partial(2\gamma)}\right)^{2m} =
\nn \\
= -\frac{1}{2}\left(\rho^2 + 
\frac{\rho^4}{8}\frac{\partial^2}{\partial\gamma^2} + 
\frac{\rho^6}{192}\frac{\partial^4}{\partial\gamma^4}
+ \ldots\right)
\ee
(note that $\sum_{m=0}^\infty \frac{x^{m+1}}{m!(m+1)!} = J_1(2x)$,
a modified Bessel function).

Using this new notation, we can rewrite (\ref{prepexpansion}) as

\be
\int_{i} dS_{DV} = \sum_{\epsilon,\bar\epsilon}
\left( \prod_{l\in\bar\epsilon} \hat D(\rho_l;\gamma_{il})\right)
\tilde I^{[i]}_{\epsilon,\bar\epsilon}(\rho_i,\gamma_{i\cdot}),
\label{prepexpansion2}
\ee
where the integrals

\be
\tilde I^{[i]}_{\epsilon,\bar\epsilon}(\rho_i,\gamma_{i\cdot}) =
\int\frac{\prod_{k\in\epsilon}(x + \gamma_{ik})}
{\prod_{l\in\bar\epsilon}(x+\gamma_{il})}
\sqrt{x-\rho_i^2}\ dx
\label{tildeintegral}
\ee
are just another representation of (\ref{integral}) with indices
$m_j$ taking only two values: $0$ (for all $j\in \epsilon$) and
$1$ (for all $j\in \bar\epsilon$).

While (\ref{tildeintegral}) is more convenient for
the general formulas like (\ref{prepexpansion2})
notation (\ref{integral}) can still be of more use in concrete 
applications (for particular $n$). For example,
in the case of {\bf $n=2$} we have just two terms in 
(\ref{prepexpansion2}): \\
${\bf n=2:}$

\be
\int_1 dS_{DV} = I^{[1]}_0(\rho_1;\gamma_{12}) +
\hat D(\rho_2;\gamma_{12}) I^{[1]}_1(\rho_1;\gamma_{12})
\label{n2a}
\ee
with

\be
I^{[1]}_0 = \tilde I^{[1]}_{\{2\};\emptyset} =
\int (x+\gamma_{12})\sqrt{x^2-\rho_1^2}dx, \nn \\
I^{[1]}_1 = \tilde I^{[1]}_{\emptyset;\{2\}} =
\int \frac{\sqrt{x^2-\rho_1^2}}{x+\gamma_{12}}dx,
\label{n2b}
\ee
and in the case of {\bf $n=3$} there are already four terms: \\
${\bf n=3:}$

\be
\int_1 dS_{DV} = I^{[1]}_{00}(\rho_1;\gamma_{12},\gamma_{13}) +
\hat D(\rho_2;\gamma_{12}) I^{[1]}_{10}(\rho_1;\gamma_{12},\gamma_{13}) +
\nn \\ +
\hat D(\rho_3;\gamma_{13}) I^{[1]}_{01}(\rho_1;\gamma_{12},\gamma_{13})
+ \hat D(\rho_2;\gamma_{12})\hat D(\rho_3;\gamma_{13}) 
I^{[1]}_{11}(\rho_1;\gamma_{12},\gamma_{13})
\label{n3a}
\ee
with

\be
I^{[1]}_{00} = \tilde I^{[1]}_{\{2,3\};\emptyset} =
\int (x+\gamma_{12})(x+\gamma_{13})\sqrt{x^2-\rho_1^2}dx, 
\nn \\
I^{[1]}_{10} = \tilde I^{[1]}_{\{3\};\{2\}} =
\int \frac{x+\gamma_{13}}{x+\gamma_{12}}\sqrt{x^2-\rho_1^2}\ dx,
\nn \\
I^{[1]}_{01} = \tilde I^{[1]}_{\{2\};\{3\}} =
\int \frac{x+\gamma_{12}}{x+\gamma_{13}}\sqrt{x^2-\rho_1^2}\ dx,
\nn \\
I^{[1]}_{11} = \tilde I^{[1]}_{\emptyset;\{2,3\}} =
\int \frac{\sqrt{x^2-\rho_1^2}}{(x+\gamma_{12})(x+\gamma_{13})}dx.
\label{n3b}
\ee

\subsection{Handling the integrals}

In order to evaluate the integrals 
$\tilde I^{[i]}_{\epsilon;\bar\epsilon}$ 
in (\ref{tildeintegral}), it is first 
necessary to transform the {\it integrands}.

First, the denominator:

\be
\prod_{l\in\bar\epsilon}\frac{1}{x+\gamma_{il}} =
\sum_{l\in\bar\epsilon}
\frac{r^{(\bar\epsilon)}(\gamma_{l\cdot})}{x+\gamma_{il}},
\label{denom}
\ee
where  

\be
r^{(\bar\epsilon)}(\gamma_{l\cdot}) \equiv
\prod_{l'\in\bar\epsilon_l}\frac{1}{\gamma_{ll'}}
\ee
(let us remind that $\bar\epsilon_l$ is obtained from $\bar\epsilon$
by throwing away the element $l\in\bar\epsilon$).

Second, for each term in the sum (\ref{denom}) we appropriately
transform the numerator. There are two rather different ways to do it,
leading to two different representations of the integral.

\subsubsection{Representation through symmetric functions}

One way is to expand the numerator in (\ref{tildeintegral}) 
for the $l$-th item of the sum (\ref{denom}) in powers of
$(x+\gamma_{il})$:

\be
\prod_{k\in\epsilon}(x + \gamma_{ik}) =
\prod_{k\in\epsilon}\left((x + \gamma_{il})+ \gamma_{lk}\right)
= \nn \\ =
\sum_{m=0}^{d(\epsilon)} (x+\gamma_{il})^{d(\epsilon)-m}
e_m^{(\epsilon)}(\gamma_{l\cdot})
= \sum_{m=-1}^{d(\epsilon)-1} (x+\gamma_{il})^{m+1}
e_{d(\epsilon)-1-m}^{(\epsilon)}(\gamma_{l\cdot}).
\ee
Here

\be
e_m^{(\epsilon)}(\gamma_{l\cdot}) =
\sum_{\stackrel{k_1<\ldots<k_m}{k_1,\ldots,k_m\in\epsilon}}
\gamma_{lk_1}\ldots\gamma_{lk_m}
\ee
are symmetric polynomials, made from the set of $\gamma_{lk}$
with given $l\in\bar\epsilon$ and all $k\in\epsilon$.
Sometime it can be more convenient to use "exclusive" notation
instead of "inclusive" one:
$e_m^{[\bar\epsilon]}(\gamma_{l\cdot}) \equiv 
e_m^{(\epsilon)}(\gamma_{l\cdot})$.
(Note that symmetric functions $r^{(\bar\epsilon)}(\gamma_{l\cdot})$
and $e_m^{(\epsilon)}(\gamma_{l\cdot})$ do not explicitly depend on $i$).

With the help of these expressions we obtain for the integral
(\ref{tildeintegral}), 

\be
I^{[i]}_{0\ldots 0}(\rho_i;\gamma_{i\cdot}) =
\sum_{m=0}^{n-1}e_{n-m-1}^{[i]}(\gamma_{i\cdot})
J_m(\rho_i;0).
\ee
for $\bar\epsilon = \emptyset$ and

\be
\tilde I^{[i]}_{\epsilon,\bar\epsilon}(\rho_i,\gamma_{i\cdot}) =
\sum_{l\in \bar\epsilon} 
r^{(\bar\epsilon)}(\gamma_{l\cdot})
\sum_{p=-1}^{d(\epsilon)-1}  
e_{d(\epsilon)-1-p}^{(\epsilon)}(\gamma_{l\cdot})
J_p(\rho_i;\gamma_{il}),
\label{intthrep}
\ee
for $\bar\epsilon\neq\emptyset$, 
where the remaining integrals depend on just two parameters:

\be
J_{p}(\rho;\gamma) \equiv \int (x+\gamma)^p\sqrt{x^2-\rho^2}dx 
= \sum_{m=0}^p \frac{p!}{m!(p-m)!}\gamma^{p-m}J_{m}(\rho;0), 
\ \ for\ p\geq 0.
\label{elimgamma}
\ee

Thus

\be
\int_i dS_{DV} = I^{[i]}_{0\ldots 0}(\rho_i;\gamma_{i\cdot}) +
\nn \\ +
\sum_{\stackrel{\epsilon,\bar\epsilon} 
{\bar\epsilon\neq\emptyset}}
\left(\prod_{l\in\bar\epsilon} \hat D(\rho_l,\gamma_{il})\right)
\left(
\sum_{l\in \bar\epsilon} 
r^{(\bar\epsilon)}(\gamma_{l\cdot})
\left[
\sum_{p=-1}^{d(\epsilon)-1}
e_{d(\epsilon)-p-1}^{(\epsilon)}(\gamma_{l\cdot})
J_p(\rho_i;\gamma_{il})
\right]\right)
\label{prepexpansion30}
\ee
The disadvantage of eq.(\ref{prepexpansion30}) is the need to
supplement it by (\ref{elimgamma}), which spoils the relatively
nice structure of (\ref{prepexpansion30}):  
the sum in the square brackets should be actually transformed to

\be
\sum_{p=-1}^{d(\epsilon)-1}
e_{d(\epsilon)-p-1}^{(\epsilon)}(\gamma_{l\cdot})
J_p(\rho_i;\gamma_{il}) = 
e_{d(\epsilon)}^{(\epsilon)}(\gamma_{l\cdot})
J_{-1}(\rho_i;\gamma_{il}) +
\sum_{m=0}^{d(\epsilon)-1}
\varepsilon_{m}^{(\epsilon)}(\gamma_{i\cdot}|\gamma_{l\cdot})
J_m(\rho_i;0)
\ee
with

\be
\varepsilon_{m}^{(\epsilon)}(\gamma_{i\cdot}|\gamma_{l\cdot})
= \sum_{p=m}^{d(\epsilon)-1} \frac{p!}{m!(p-m)!}
e_{d(\epsilon)-p-1}^{(\epsilon)}(\gamma_{l\cdot})\gamma_{il}^{p-m}
\ee
or

\be
\varepsilon_{d(\epsilon)-m-1}^{(\epsilon)}(\gamma_{i\cdot}|\gamma_{l\cdot})
= \sum_{p=0}^{m} \frac{(d(\epsilon)-1-p)!}{(d(\epsilon)-1-m)!(m-p)!}
e_{p}^{(\epsilon)}(\gamma_{l\cdot})\gamma_{il}^{m-p}
\label{varepvse}
\ee
However, once the need to introduce the $\varepsilon$-coefficients is
accepted, it is better to do it earlier, handling the numerator
in (\ref{tildeintegral}) in a different way from the very beginning.

\subsubsection{Dividing the polynomials}

Another way to deal with the numerator in the $l$-th item of
the sum (\ref{denom}) is just to {\it divide} it by $(x+\gamma_{il})$, 
obtaining the ratio polynomial of degree $d(\epsilon)-1$ and the
residue:\footnote{Note that symmetric functions $e^{(\epsilon)}$ 
at the l.h.s. of (\ref{varepdef}) have
the arguments $\gamma_{ik}$, $k\in\epsilon$, while
in the previous subsection 2.2.1 their arguments  were rather $\gamma_{lk}$
with $l\in\bar\epsilon$.
Accordingly, $\gamma_{i\cdot}$
can seem to be a better choice for the argument of $\varepsilon$-functions, if
one looks at (\ref{varepdef}), and the same can seem about
$\gamma_{l\cdot}$ if one looks at (\ref{varepvse}). 
In fact, the truth is somewhere in between; $\varepsilon_m$
with low $m$ are closer to $e(\gamma_{l\cdot})$, while those
with high $m$ are closer to $e(\gamma_{i\cdot})$.
We use a somewhat heavy notation, with both types of arguments
included. It certainly deserves to be improved.
}

\be
\prod_{k\in\epsilon}(x+\gamma_{ik}) 
= \sum_{p=0}^{d(\epsilon)} e^{(\epsilon)}_{d(\epsilon)-p}(\gamma_{i\cdot})
x^p =
\varepsilon^{(\epsilon)}_{-1}(\gamma_{i\cdot}|\gamma_{l\cdot}) +
(x+\gamma_{il})\sum_{m=0}^{d(\epsilon)-1}
\varepsilon^{(\epsilon)}_{m}(\gamma_{i\cdot}|\gamma_{l\cdot})x^m
\label{varepdef}
\ee
The functions $\varepsilon^{(\epsilon)}_m$ are expressed through
$e^{(\epsilon)}_{p}(\gamma_{i\cdot})$ by recurrent relations,

\be
\varepsilon^{(\epsilon)}_{m-1} + \gamma_{il}\varepsilon^{(\epsilon)}_m =
e^{(\epsilon)}_{d(\epsilon)-m}(\gamma_{i\cdot}),
\ee
so that

\be
\varepsilon^{(\epsilon)}_{d(\epsilon)-1} = 
e^{(\epsilon)}_{0}(\gamma_{i\cdot}) = 1,
\nn \\
\varepsilon^{(\epsilon)}_{d(\epsilon)-2} = 
e^{(\epsilon)}_{1}(\gamma_{i\cdot}) -
\gamma_{il} e^{(\epsilon)}_{0}(\gamma_{i\cdot}) 
= - \gamma_{il} + \sum_{k\in\epsilon}\gamma_{ik}, 
\nn \\
\ldots
\nn \\
\varepsilon^{(\epsilon)}_{0} = \frac{\tilde\Delta^{(\epsilon)}_i
-\tilde\Delta^{(\epsilon)}_l}{\gamma_{il}},
\nn \\
\varepsilon^{(\epsilon)}_{-1} = e^{(\epsilon)}_{d(\epsilon)}(\gamma_{l\cdot})
\equiv \tilde \Delta^{(\epsilon)}_l 
\ee
The last lines here are expressed through

\be
\tilde \Delta^{(\epsilon)}_l \equiv \tilde \Delta^{[i,\bar\epsilon]}_l
\equiv \prod_{k\in\epsilon}\gamma_{lk}.
\ee
For $\varepsilon = \emptyset$ we skip the superscript:

\be
\tilde\Delta_i \equiv \tilde\Delta^{[i]}_i = \prod_{j\neq i}\gamma_{ij}.
\ee
Tilde means that the products are made from $\gamma$'s, notation without
tildes are reserved to denote the same quantities, but made from $\alpha$'s.

In terms of $\varepsilon$ functions we obtain instead of (\ref{intthrep})

\be
\tilde I^{[i]}_{\epsilon,\bar\epsilon}(\rho_i,\gamma_{i\cdot}) =
\varepsilon^{(\epsilon)}_{-1}(\gamma_{i\cdot}|\gamma_{l\cdot})
J_{-1}(\rho_i;\gamma_{il}) +
\sum_{m=0}^{d(\epsilon)-1}
\varepsilon_{m}^{(\epsilon)}(\gamma_{i\cdot}|\gamma_{l\cdot})
J_m(\rho_i;0)
\ee
for $\bar\epsilon\neq\emptyset$, 
and, finally

\be
\int_i dS_{DV} = I^{[i]}_{0\ldots 0}(\rho_i;\gamma_{i\cdot}) +
\nn \\ +
\sum_{\stackrel{\epsilon,\bar\epsilon} 
{\bar\epsilon\neq\emptyset}}
\left(\prod_{l\in\bar\epsilon} \hat D(\rho_l,\gamma_{il})\right)
\left(
\sum_{l\in \bar\epsilon} 
r^{(\bar\epsilon)}(\gamma_{l\cdot})
\left[
\varepsilon^{(\epsilon)}_{-1}(\gamma_{i\cdot}|\gamma_{l\cdot})
J_{-1}(\rho_i;\gamma_{il}) +
\sum_{m=0}^{d(\epsilon)-1}
\varepsilon_{m}^{(\epsilon)}(\gamma_{i\cdot}|\gamma_{l\cdot})
J_m(\rho_i;0)
\right]\right)
\label{prepexpansion3}
\ee
The price for obtaining a representation through the simple integrals
$J_m(\rho_i;0)$ with vanishing second argument is the appearance of
somewhat sophisticated coefficients $\varepsilon(\gamma)$. 
The $\gamma$-dependence can {\it not} be eliminated from the integral
$J_{-1}(\rho;\gamma)$, but of course it can be easily evaluated,
see the next subsection 2.3.

\subsubsection{Examples of $n=2$ and $n=3$}

The simple example of integrals in (\ref{n2b}) and (\ref{n3b}) are 
in fact not sensitive to all these complications and are easily
expressed through $J_m$-integrals.
\\
${\bf n=2:}$

\be
I^{(1)}_{0}(\rho_1;\gamma_{12})  = J_1(\rho_1;0) + \gamma_{12}J_0(\rho_1;0), 
\nn \\
I^{(1)}_{1}(\rho_1;\gamma_{12}) = J_{-1}(\rho_1;\gamma_{12}), 
\label{n2c}
\ee
and \\
${\bf n=3:}$

\be
I^{(1)}_{00}(\rho_1;\gamma_{12},\gamma_{13}) = J_2(\rho_1;0) +
(\gamma_{12}+\gamma_{13})J_1(\rho_1;0) + 
\gamma_{12}\gamma_{13}J_0(\rho_1;0), \nn \\
I^{(1)}_{10}(\rho_1;\gamma_{12},\gamma_{13}) = J_0(\rho_1;0) +
\gamma_{23}J_{-1}(\rho_1;\gamma_{12}), \nn \\
I^{(1)}_{01}(\rho_1;\gamma_{12},\gamma_{13}) = J_0(\rho_1;0) -
\gamma_{23}J_{-1}(\rho_1;\gamma_{13}), \nn \\
I^{(1)}_{11}(\rho_1;\gamma_{12},\gamma_{13}) =
\frac{1}{\gamma_{23}}\left[J_{-1}(\rho_1;\gamma_{12})-
J_{-1}(\rho_1;\gamma_{13})\right]
\label{n3c}
\ee

\subsection{$J_m$-integrals and their periods}

These elementary integrals are given by somewhat heavy-looking
formulas:\footnote{
For example,
$$
J_0 = \rho^2 \int \sinh^2 t dt = 
\frac{\rho^2}{4} \left(\sinh 2t - 2t\right),
$$
$$ 
J_1 = \rho^3 \int \cosh t\sinh^2 t dt = 
\frac{\rho^3}{12} \left(\sinh 3t - 3 \sinh t\right),
$$
$$ 
J_2 = \rho^4 \int \cosh^2 t\sinh^2 t dt = 
\frac{\rho^4}{32} \left(\sinh 4t - 4t\right),
$$
$$ 
\ldots
$$
}

\be
J_{2m}(\rho;0) = 
c_{m+1}\rho^{2m+2}t + 
\rho^{2m+2}\sum_{j=1}^{m+1}c_{m+1}^{(j)}\sinh 2jt, \ m\geq 0, \nn\\
J_{2m-1}(\rho;0) = 
\rho^{2m+1}\sum_{j=0}^{m}c_{m+1/2}^{(j+1/2)}\sinh (2j+1)t, \ m > 0, \nn \\
J_{-1}(\rho;\gamma) = 
\sqrt{x^2-\rho^2} - \gamma t +
\sqrt{\gamma^2-\rho^2}
\log\frac{\rho z + \gamma - \sqrt{\gamma^2-\rho^2}}
{\rho z + \gamma + \sqrt{\gamma^2-\rho^2}},
\ee
where

\be
c_k^{(l)} \equiv c_k^{[k-l]} \equiv
2^{1-2k}\frac{2l^2-k}{l}\cdot\frac{(2k-2)!}{(k-l)!(k+l)!}, \nn \\
c_k^{(k)} = c_k^{[0]} = \frac{1}{2^{2k}k},
c_k^{(k-1)} = c_k^{[1]} = \frac{2(k-2)}{2^{2k}(k-1)},
c_k^{(k-2)} = c_k^{[2]} = \frac{2k^2-9k + 8}{2^{2k}(k-2)},\ \ldots, \\ \nn
as\ l\rightarrow 0,\ c_k^{(l)} \sim \frac{c_k}{l}, 
\label{cexa}
\ee
and

\be
x=\rho \cosh t,\ \ \ z= e^t = \frac{x + \sqrt{x^2-\rho^2}}{\rho}.
\ee

Actually, we are interested in two types of integration paths:

$A_i$: $x$ from $\rho_i$ to $-\rho_i$,
i.e. $t$ from $0$ to $i\pi$ and $z$ from $1$ to $-1$, and

$B_i$: $x$ from $\rho_i$ to $\Lambda_i \equiv
\Lambda_{(\gamma_i)} \equiv \Lambda-\gamma_i$,
i.e. $z$ from $1$ to $Z_i/\rho_i$, where

\be
Z_{(\gamma)}(\rho) = \Lambda_{(\gamma)} + \sqrt{\Lambda_{(\gamma)}^2-\rho^2} =
2\Lambda_{(\gamma)} + 
\sum_{m=1}^\infty c_m\frac{\rho^{2m}}{\Lambda_{(\gamma)}^{2m-1}},
\ee
and $t$ from $0$ to $\log Z_{i}/\rho_i$.

We denote the integrals $J_m$ along these paths by $J^A_m$ and $J^B_m$ .
The $A_i$ and $B_i$ periods of $dS_{DV}$, i.e. $-S_i/g_{n+1}$
and $\Pi_i/g_{n+1}$
are then obtained by substituting the corresponding integrals,
into (\ref{prepexpansion}) and multiplying by an extra factor of $2$.
Note, that all the operators $\hat D(\gamma_l;\gamma_{il})$ in 
(\ref{prepexpansion3}) commute with $\Lambda_i$ and $Z_i$:
derivatives act on $\gamma_l$, {\it not} on $\gamma_i$,

For the integrals along the $A$ contour we obtain simple
expressions:

\be
J^A_{2m}(\rho;0) = i\pi c_{m+1} \rho^{2m+2}, \ m\geq 0,
\nn \\
J^A_{2m-1}(\rho;0) = 0, \ \ m>0, \nn \\
J^A_{-1}(\rho,\gamma) = -i\pi (\gamma - \sqrt{\gamma^2-\rho^2})
= i\pi \sum_{k=1}^\infty c_k\frac{\rho^{2k}}{\gamma^{2k-1}}
= \nn \\ =
-i\pi\left(\frac{\rho^2}{2\gamma} + \frac{\rho^4}{8\gamma^3}
+ \frac{\rho^6}{16\gamma^5}
+ \ldots\right)
\label{JA}
\ee
but those for the $B$ contour are more sophisticated
(see (\ref{cexa}) for the definition of $c_k^{[l]}$):

\be
J^B_{2m}(\rho;Z) = 
\frac{i}{\pi}J^A_{2m}(\rho;0) \log\frac{\rho}{2\Lambda}
+ \frac{1}{2}\sum_{j=0}^{m} c_{m+1}^{[j]}
\rho^{2j}Z^{2m+2-2j} 
+ O(\Lambda^{-1}),
\nn \\
J^B_{2m-1}(\rho;Z) = 
\frac{1}{2}\sum_{j=0}^{m} c_{m+1/2}^{[j]}\rho^{2j}Z^{2m+1-2j} 
+ O(\Lambda^{-1}), \nn \\
J^B_{-1}(\rho,\gamma|\Lambda_{(\gamma')}) = \Lambda_{(\gamma')} -
\gamma\log\frac{2\Lambda}{\rho} + \sqrt{\gamma^2-\rho^2}L(\rho;\gamma) =
\nn \\ =
\Lambda_{(\gamma')} +
\gamma\log\frac{\gamma}{\Lambda}
+\frac{i}{\pi} J^A_{-1}(\rho,\gamma)\log\frac{\rho}{2\gamma} +
\sum_{k,l=0}^\infty \frac{k+2}{k+1}c_lc_{k+2}
\frac{\rho^{2k+2l+2}}{\gamma^{2k+2l+1}} = \nn \\ =
\Lambda_{(\gamma')} +
\gamma\log\frac{\gamma}{\Lambda}
+\frac{i}{\pi} J^A_{-1}\log\frac{\rho}{2\gamma} -
\frac{\rho^2}{4\gamma} + \frac{\rho^4}{32\gamma^3} +
\frac{5\rho^6}{192\gamma^5} + \frac{59\rho^8}{3\cdot 1024\gamma^7}
+ \ldots
\label{JB}
\ee
Here 

\be
\frac{\partial L(\rho,\gamma)}{\partial\gamma} =
\frac{1}{\sqrt{\gamma^2-\rho^2}} = 
\sum_{k=0}^\infty \frac{\Gamma(k+1/2)}{k!\Gamma(1/2)}
\frac{\rho^{2k}}{\gamma^{2k-1}} =
-2\sum_{k=1}^\infty (k+1)c_{k+1} \frac{\rho^{2k}}{\gamma^{2k-1}}, \nn \\
L(\rho;\gamma) = \log
\frac{\rho+\gamma+\sqrt{\gamma^2-\rho^2}}{\rho+\gamma-\sqrt{\gamma^2-\rho^2}}
= \log\frac{2\gamma}{\rho} + 
\sum_{k=1}^\infty \frac{(k+1)c_{k+1}}{k} \frac{\rho^{2k}}{\gamma^{2k}},\nn \\
\sqrt{\gamma^2-\rho^2}L(\rho;\gamma)  =
\sqrt{\gamma^2-\rho^2}\log\frac{2\gamma}{\rho} +
\sum_{k,l=0}^\infty \frac{(k+2)c_{k+2}c_l}{k} 
\frac{\rho^{2k+2l+2}}{\gamma^{2k+2l+1}}
\ee

\section{The first three terms of the $\rho^2$-expansion}

\subsection{Truncations and $S$-integrals}

Let us now pick up the first terms of expansion in powers of $\rho^2$
in (\ref{prepexpansion3}). Since $S=O(\rho^2)$, for the purpose of
calculating ${\cal F}_3(S|\alpha)$ it is enough to keep the first three
terms, of orders $\rho^0$, $\rho^2$ and $\rho^4$.
This implies that $d(\bar\epsilon)\leq 2$ and 
leaves only three options for $\bar\epsilon$:
$\bar\epsilon = \emptyset$, $\bar\epsilon = \{l\}$ and
$\bar\epsilon = \{l_1,l_2\}$, so that (\ref{prepexpansion3}) is
truncated to

\be
\int_i dS_{DV} = 
\sum_{m=0}^{n-1} e^{[i]}_{n-m-1}(\gamma_{i\cdot}) J_m(\rho_i;0) - 
\nn \\
- \frac{1}{2} \sum_{l\neq i}  
\left(
\rho_l^2 + \frac{\rho_l^4}{8}\frac{\partial^2}{\partial\gamma_l^2}
\right)
\left[
\varepsilon^{[il]}_{-1}(\gamma_{i\cdot}|\gamma_{l\cdot})
J_{-1}(\rho_i;\gamma_{il}) +
\sum_{m=0}^{n-3}
\varepsilon^{[il]}_{m}(\gamma_{i\cdot}|\gamma_{l\cdot})
J_m(\rho_i;0)
\right] + 
\nn \\
+ \frac{1}{4}\sum_{\stackrel{l_1<l_2}{l_1,l_2\neq i}} 
\frac{\rho_{l_1}^2\rho_{l_2}^2}{\gamma_{l_1l_2}}
\left[       
\varepsilon^{[il_1l_2]}_{-1}(\gamma_{i\cdot}|\gamma_{l_1\cdot})
J_{-1}(\rho_i;\gamma_{il_1}) -
\varepsilon^{[il_1l_2]}_{-1}(\gamma_{i\cdot}|\gamma_{l_2\cdot})
J_{-1}(\rho_i;\gamma_{il_2}) +  \phantom{\sum_{m=0}^{n-4}}
\right. \nn \\ \left. +
\sum_{m=0}^{n-4}\left(
\varepsilon^{[il_1l_2]}_m(\gamma_{i\cdot}|\gamma_{l_1\cdot})
J_m(\rho_i;0) -
\varepsilon^{[il_1l_2]}_m(\gamma_{i\cdot}|\gamma_{l_2\cdot})
J_m(\rho_i;0)
\right)
\right]
\label{truncated1}
\ee

The sums over $m$ could also be truncated, if there were no
$\Lambda_i$-dependent terms in $J^B_m$. However, for $A$-integrals
there is no such problem, and only terms with $m\leq 2$ 
in the first line and with $m=0$ in the second one can 
contribute, because $J_m^A = O(\rho^{m+2})$.
Moreover, since also $J_{-1}^A = O(\rho^{2})$ (note, that this is
{\it not} true for $J_{-1}^B$, even if the $\Lambda_i$-dependent
terms were excluded), the last term in (\ref{truncated1}) is
also irrelevant, and we get for the $A_i$-period $S_i$ of $dS_{DV}$:

\be
-\frac{1}{2\pi i g_{n+1}}S_i = -\frac{1}{2}
e^{[i]}_{n-1}(\gamma_{i\cdot})\rho_i^2 - 
\frac{1}{8}e^{[i]}_{n-3}(\gamma_{i\cdot})\rho_i^4 +
\nn\\  
+ \frac{1}{4}\sum_{l\neq i}\rho_i^2\rho_l^2
\left(
\frac{\varepsilon^{[il]}_{-1}(\gamma_{i\cdot}|\gamma_{l\cdot})}{\gamma_{li}}
+ \varepsilon^{[il]}_{0}(\gamma_{i\cdot}|\gamma_{l\cdot})
\right)
+ O(\rho^6),
\label{Sgen}
\ee
so that for $\hat S_i \equiv \frac{S_i}{i\pi g_{n+1}}$ and

\be
e^{[i]}_{n-1}(\gamma_{i\cdot}) = 
\prod_{j\neq i}\gamma_{ij} \equiv \tilde\Delta_i, 
\nn \\
e^{[i]}_{n-3}(\gamma_{i\cdot}) = \tilde\Delta_i
\sum_{\stackrel{j<k}{j,k\neq i}}\frac{1}{\gamma_{ij}\gamma_{ik}}, 
\nn \\
\varepsilon^{[il]}_{-1}(\gamma_{i\cdot}|\gamma_{l\cdot}) =
\prod_{k\neq i,l}\gamma_{lk} \equiv \tilde\Delta_l^{[il]} =
\frac{\tilde \Delta_l}{\gamma_{li}}, 
\nn \\ 
\varepsilon^{[il]}_{0}(\gamma_{i\cdot}|\gamma_{l\cdot}) =
\frac{\tilde\Delta^{[il]}_i - \tilde\Delta^{[il]}_l}{\gamma_{il}}
= \frac{\tilde\Delta_i + \tilde\Delta_l}{\gamma_{il}^2}
\ee
we get:

\be
\hat S_i = 
\rho_i^2\tilde \Delta_i + 
\frac{1}{4}\rho_i^4\tilde \Delta_i
\sum_{\stackrel{j<k}{j,k\neq i}}\frac{1}{\gamma_{ij}\gamma_{ik}}
- \frac{1}{2}\rho_i^2\sum_{l\neq i}
\left(
-\frac{\tilde\Delta_l}{\gamma_{li}^2} + 
\frac{\tilde\Delta_i + \tilde\Delta_l}{\gamma_{il}^2}
\right)
= \nn \\ =
\rho_i^2\tilde \Delta_i
\left(
1 + \frac{1}{4}\rho_i^2
\sum_{\stackrel{j<k}{j,k\neq i}}\frac{1}{\gamma_{ij}\gamma_{ik}}
- \frac{1}{2}\sum_{j\neq i}\frac{\rho_j^2}{\gamma_{ij}^2}
+ O(\rho^4)
\right) 
\label{Srho}
\ee

and
\be
\rho_i^2 = \frac{\hat S_i}{\tilde\Delta_i} \left(1 +
\sum_{j=1}^n \zeta_{ij}(\gamma)\hat S_j + O(S^2)
\right),
\nn \\
\zeta_{ii} = -\frac{1}{4\tilde\Delta_i}
\sum_{\stackrel{j<k}{j,k\neq i}}\frac{1}{\gamma_{ij}\gamma_{ik}},
\nn \\
\zeta_{ij} = -\frac{1}{\gamma_{ij}^2\tilde\Delta_j},\ \ j\neq i. 
\label{zeta}
\ee
Remarkably, due to peculiar relation\footnote{
The combination 
$S\log S - S = 
\frac{1}{2}\partial_S\left[S^2(\log S - \frac{3}{2})\right]$ 
is characterized by the fact that its $S$-derivative is exactly
$\log S$.
}

\be
\hat S_i\log(\Delta^{[i]}\rho_i^2) - (\Delta^{[i]}\rho_i^2) = 
\hat S_i\log \hat S_i - \hat S_i + O(S^3),
\label{slog-s}
\ee
the coefficients $\zeta$ from (\ref{zeta}) do not show up in explicit
expression for ${\cal F}_3(S|\alpha)$, though the entire eq.(\ref{Sgen})
is needed to handle the logarithmic contributions to the prepotential.

\subsection{$\Pi$-integral}

Let us now proceed to evaluation of the $\Pi_i$ integral:

\be
\frac{\Pi_i}{2g_{n+1}} =
\sum_{m\geq 0}^{n-1} e^{[i]}_{n-m-1}(\gamma_{i\cdot}) J^B_m(\rho_i;0) - 
\nn \\
- \frac{1}{2} \sum_{l\neq i} 
\left(\rho_l^2 + \frac{\rho_l^4}{8}\frac{\partial^2}{\partial\gamma_l^2}
\right)
\left[
\varepsilon^{[il]}_{-1}(\gamma_{i\cdot}|\gamma_{l\cdot})
J^B_{-1}(\rho_i;\gamma_{il}) +
\sum_{m=0}^{n-3}
\varepsilon^{[il]}_{m}(\gamma_{i\cdot}|\gamma_{l\cdot})
J^B_m(\rho_i;0)
\right] + 
\nn \\
+ \frac{1}{4}\sum_{\stackrel{l_1<l_2}{l_1,l_2\neq i}}
\frac{\rho_{l_1}^2\rho_{l_2}^2}{\gamma_{l_1l_2}}
\left[
\varepsilon^{[il_1l_2]}_{-1}(\gamma_{i\cdot}|\gamma_{l_1\cdot})
J^B_{-1}(\rho_i;\gamma_{il_1}) -
\varepsilon^{[il_1l_2]}_{-1}(\gamma_{i\cdot}|\gamma_{l_2\cdot})
J^B_{-1}(\rho_i;\gamma_{il_2}) 
\right. \nn \\ \left. +
\sum_{m=0}^{n-4}\left(
\varepsilon^{[il_1l_2]}_m(\gamma_{i\cdot}|\gamma_{l_1\cdot})
J^B_m(\rho_i;0) -
\varepsilon^{[il_1l_2]}_m(\gamma_{i\cdot}|\gamma_{l_2\cdot})
J^B_m(\rho_i;0)
\right)
\right]
+ \nn \\ + O(\rho^6)
\label{Pi1}
\ee
where expressions for $J^B_m$  should be substituted from (\ref{JB}).
After that the r.h.s. consists of the contributions of three essentially
different types: containing $\Lambda_i$, containing logarithms
(of any kind of arguments, $\Lambda$, $\rho$ and $\gamma$),
and power-like terms, proportional to $\rho^2$ and $\rho^4$
with $\gamma$-dependent coefficients. In the rest of this
{\it sub}section we analyze these three types of terms separately, 
giving some details (one can easily understand how things work,
looking through explicit examples in Appendix B).

\subsubsection{$\Lambda$-dependent terms}

The $\Lambda_i$-dependent terms are:

\be
\frac{1}{2}\sum_{m=0}^{n-1} e^{[i]}_{n-m-1}(\gamma_{i\cdot})
\left[c_{m/2+1}^{[0]}(2\Lambda_i)^{m+2}
\left(1-\frac{(m+2)\rho_i^2}{4\Lambda_i^2}
+ \frac{(m-1)(m+2)\rho_i^4}{32\Lambda_i^4}
\right) +
\right.\nn \\ \left.
+ \theta(m\geq 1)c_{m/2+1}^{[1]}\rho_i^2(2\Lambda_i)^m 
\left(1-\frac{m\rho_i^2}{4\Lambda_i^2}\right)
+ \theta(m\geq 3)c_{m/2+1}^{[2]}\rho_i^4(2\Lambda)^{m-2} 
\right] -
\nn \\ 
- \frac{1}{4} \sum_{l\neq i}\rho_l^2 
\sum_{m=-1}^{n-3} 
\varepsilon^{[il]}_m(\gamma_{i\cdot}|\gamma_{l\cdot})
\left[c_{m/2+1}^{[0]}(2\Lambda_i)^{m+2}
\left(1-\frac{(m+2)\rho_i^2}{4\Lambda_i^2}
\right) 
+ \theta(m\geq 1)c_{m/2+1}^{[1]}\rho_i^2(2\Lambda_i)^m 
\right]
\nn \\ +
\frac{1}{8}\sum_{\stackrel{l_1<l_2}{l_1,l_2\neq i}}
\frac{\rho_{l_1}^2\rho_{l_2}^2}{\gamma_{l_1l_2}}
\frac{1}{2}\sum_{m=0}^{n-4}
\left[
\varepsilon^{[il_1l_2]}_m(\gamma_{i\cdot}|\gamma_{l\cdot}) -
\varepsilon^{[il_1l_2]}_m(\gamma_{i\cdot}|\gamma_{l\cdot})
\right]
c_{m/2+1}^{[0]}(2\Lambda_i)^{m+2}
+ \nn \\ + O(\rho^6)
\label{Lambdaterms}
\ee
The relevant coefficients $c_k^{[l]}$ are listed in (\ref{cexa}),

\be
\frac{1}{2}c_{m/2+1}^{[0]} = \frac{1}{2^{m+2}(m+2)},\ \
\frac{1}{2}c_{m/2+1}^{[1]} = \frac{1}{2^{m+2}}\frac{m-2}{m},\ \
\frac{1}{2}c_{m/2+1}^{[2]} = \frac{1}{2^{m+2}}\frac{m^2-5m+2}{2(m-2)}.
\ee
An extra coefficient $1/2$ as compared to (\ref{Pi1}) appears in these
formulas because $J^B_m \sim \frac{1}{2}Z_i^{m+2}$ for large $Z_i$.
This is not true for $J^B_{-1}$, but instead

\be
\frac{1}{2}c_{m/2+1}^{[0]}(2\Lambda_i)^{m+2} = 
\frac{1}{m+2}\Lambda_i^{m+2}
\ee 
is exactly $\Lambda_i$ for $m=-1$, so the term with $m=-1$ is
naturally included into the sum at the r.h.s. of (\ref{Lambdaterms}).
 We have introduced the Heaviside step function: 
 $\theta(m\geq k)=0$ for $m<k$ and $\theta(m\geq k)=1$ for $m\geq k$.

It is especially easy to see what happens to the $\rho$-independent terms.
By definition of symmetric polynomial $e_m(\gamma_{i\cdot})$,

\be
\sum_{m=0}^{n-1} e^{[i]}_{n-m-1}(\gamma_{i\cdot})\Lambda_i^m
= \prod_{j\neq i}^n (\Lambda_i + \gamma_{ij}) 
= \prod_{j\neq i}^n(\Lambda - \gamma_j) =
\frac{1}{\Lambda_i}\tilde P_n(\Lambda)
\label{111}
\ee
with

\be
\tilde P_n(x) = \prod_{j=1}^n(x-\gamma_j).
\ee 
It remains to multiply both sides of (\ref{111}) by an extra 
$\Lambda_i = \Lambda-\gamma_i$ and integrate over $\Lambda_i$ 
to get for the leading term in (\ref{Lambdaterms}):

\be
\sum_{m=0}^{n-1} e^{[i]}_{n-m-1}(\gamma_{i\cdot})
\frac{(2\Lambda_i)^{m+2}}{2^{m+2}(m+2)} =
{\tilde W_{n+1}(\Lambda) - \tilde W_{n+1}(\gamma_i)}
\label{112}
\ee
with $\tilde W_{n+1}'(x) = \tilde P_n(x)$.
To check this reasoning it is enough to take the $\Lambda$-derivative
of eq.(\ref{112}): it obviously gives eq.(\ref{111}) times $\Lambda_i$.
If one proceeds in this way, the integration constant at the r.h.s. of
(\ref{112}) is equal to $-\tilde W_{n+1}(\gamma_i)$ in order to make
the r.h.s. vanishing at $\Lambda=\gamma_i$.

The role of the terms with both $\Lambda_i$ and 
$\rho^2$-dependence is to change $\tilde W_{n+1}(x)$
with $\gamma$-dependent coefficients for $W_{n+1}(x)$ with those,
made from $\alpha$'s. This is a small correction, since
$\sigma_i = \gamma_i - \alpha_i = O(\rho^2)$ and thus
$\tilde P_n(x) - P_n(x) = O(\rho^2)$ and
$\tilde W_{n+1}(x) - W_{n+1}(x) = O(\rho^2)$. 
Note, that for a given power of $\Lambda_i$
there is only a single power of $\rho^2$ that contributes:
no infinite series in powers of $\rho^2$ are involved into the
relation between $\tilde W_{n+1}(x)$ and $W_{n+1}(x)$. 
It happens so because this relation actually contains $\rho$ only 
in the form of the sum rules (\ref{sumrules2}). Just for the
same reason one does not need to know higher corrections to
(\ref{sigmaS}), though at the first glance they could seem to be important:
the sum rules (\ref{sumrules2}) can be applied directly.
The simplest way to check all these statements is again to compare
the $\Lambda$-derivative of (\ref{Lambdaterms}) with that
of $W_{n+1}(\Lambda)$, which is equal to

\be
W'_{n+1}(\Lambda) = \prod_{j=1}^n(\Lambda - \alpha_j) =
\prod_{j=1}^n(\Lambda - \alpha_i+\alpha_{ij}) =
(\Lambda_i + \sigma_i)\prod_{j\neq i}
\left[(\Lambda_i+\sigma_i)+\alpha_{ij}\right] = 
\nn \\ =
\sum_{m=0}^{n-1} e^{[i]}_{n-m-1}(\alpha_{i\cdot})
(\Lambda_i+\sigma_i)^{m+1} =
\nn \\ =
\sum_{m=0}^{n-1} e^{[i]}_{n-m-1}(\alpha_{i\cdot})\Lambda_i^{m+1}
+ \sigma_i\sum_{m=0}^{n-1} (m+1)e^{[i]}_{n-m-1}(\alpha_{i\cdot})\Lambda_i^m
+ \nn \\ + \ \ \ \
\frac{\sigma_i^2}{2}
\sum_{m=0}^{n-1} m(m+1)e^{[i]}_{n-m-1}(\alpha_{i\cdot})\Lambda_i^{m-1}
+ O(\rho^6)
\ee
For example, in the order $O(\rho^4)$ we have:

\be
\sum_{m=0}^{n-1} e^{[i]}_{n-m-1}(\gamma_{i\cdot})
\left(\Lambda_i^{m+1} - \frac{\rho_i^2\Lambda_i^{m-1}}{2}\theta(m\geq 1)\right)
- \frac{1}{2}\sum_{l\neq i}\rho_l^2 \sum_{m=0}^{n-2}
\varepsilon^{[il]}_{m-1}(\gamma_{i\cdot}|\gamma_{l\cdot})\Lambda_i^m =
\nn \\ =
\sum_{m=0}^{n-1} e^{[i]}_{n-m-1}(\alpha_{i\cdot})\Lambda_i^{m+1}
+ \sigma_i\sum_{m=0}^{n-1} (m+1)e^{[i]}_{n-m-1}(\alpha_{i\cdot})\Lambda_i^m
+ O(\rho^4)
\ee
Collecting contributions with the same powers of $\Lambda_i$, we get:

\be
e^{[i]}_0(\gamma_{i\cdot}) = e^{[i]}_0(\alpha_{i\cdot}), \ \
i.e. \ 1=1; 
\nn \\ \nn \\
e^{[i]}_1(\gamma_{i\cdot}) = e^{[i]}_1(\alpha_{i\cdot}) 
+ n\sigma_ie^{[i]}_0(\alpha_{i\cdot}), \ \
i.e. \ 
\sum_{j\neq i} \gamma_{ij} = \sum_{j\neq i}\alpha_{ij} + n\sigma_i; 
\nn \\ \nn \\
e^{[i]}_2(\gamma_{i\cdot}) - 
\frac{\rho_i^2e^{[i]}_0(\gamma_{i\cdot})}{2}
- \frac{1}{2}\sum_{l\neq i}\rho_l^2
\varepsilon^{[il]}_{n-3}(\gamma_{i\cdot}|\gamma_{l\cdot})
= e^{[i]}_2(\alpha_{i\cdot}) + (n-1)e^{[i]}_1(\alpha_{i\cdot})\sigma_i, 
\nn \\
i.e. \ 
\sum_{\stackrel{j<k}{j,k\neq i}}\gamma_{ij}\gamma_{ik} -
\frac{1}{2}\sum_{j=1}^n \rho_j^2 =
\sum_{\stackrel{j<k}{j,k\neq i}}\alpha_{ij}\alpha_{ik}
+ (n-1)\sigma_i\sum_{j\neq i}\alpha_{ij};
\nn \\
\ldots
\label{115}
\ee
The next step is to express 
$e^{[i]}_m(\gamma_{i\cdot})$ through $\gamma_i$ and $e_k(\gamma)$
  (or $p_k(\gamma)$)
and $e^{[i]}_m(\alpha_{i\cdot})$ through $\alpha_i$ and $e_k(\alpha)$
 (or $p_k(\gamma)$).
For example,

\be
e^{[i]}_2(\alpha_{i\cdot}) = 
\sum_{\stackrel{j<k}{j,k\neq i}}\alpha_{ij}\alpha_{ik} = \nn \\
= \frac{1}{2}\left(\sum_{j=1}^n\alpha_j\right)^2 -
\frac{1}{2}\sum_{j=1}^n\alpha_j^2 - (n-1)\alpha_i \sum_{n=1}^j\alpha_j
+ \frac{n(n-1)}{2}\alpha_i^2 = \nn \\ =
\frac{p_1^2(\alpha)}{2} - \frac{p_2(\alpha)}{2} -
(n-1)\alpha_i p_1(\alpha) + \frac{n(n-1)}{2}\alpha_i^2
\ee
After that eqs.(\ref{115}) are easily reduced to the sum rules
(\ref{sumrules}), relating $e_m(\alpha)$ and $e_m(\gamma)$.
We do not go in further details of this calculation. 
A more concise way to make and present it in full generality
remains to be found.

Of more interest for us below are a few terms with $-1\leq m \leq 2$,
which do not depend on $\Lambda_i$ at all.
They drop out after when the $\Lambda$-derivative is taken and do
not contribute to above calculation. Instead they will
contribute to ${\cal F}_3$. There are exactly three such terms,
coming from integrals $J^B_2$ and $J^B_0$,  the relevant
coefficients are $c_2^{[0]} = c_2^{(2)} =
1/32$ and $c_1^{[0]}=c_1^{(1)} = 1/4$: 

\be
-\frac{1}{4}e^{[i]}_{n-1}(\gamma_{i\cdot})\rho_i^2
+ \frac{1}{32}e^{[i]}_{n-3}(\gamma_{i\cdot})\rho_i^4
+ \frac{1}{8}\rho_i^2\sum_{l\neq i}
\varepsilon^{[il]}_0(\gamma_{i\cdot}|\gamma_{l\cdot})\rho_l^2
\label{lambdaindep}
\ee

\subsubsection{Logarithmic terms}

Logarithmic terms are easy to collect.
All such terms, coming from $J_{2m}^B$ with $m\geq 0$
are equal to the corresponding $J^A_{2m}$, multiplied by
$\log\frac{\rho_i}{2\Lambda}$, which commutes with
$\hat D(\rho_i;\gamma_{il})$. Thus they combine
altogether to provide 
$\frac{i}{\pi}(S_i/2g_{n+1}) = \frac{1}{2}\hat S_i$,
multiplied by the same logarithm. 
One should take care only of the contributions from 

\be
J^B_{-1} = \frac{i}{\pi}J^A_{-1}\left(\log \frac{\rho}{2\Lambda} -
\log\frac{\gamma}{\Lambda}\right) + \gamma\log\frac{\gamma}{\Lambda} 
+ \ {non-logarithmic}\ terms = 
\nn \\ =
\frac{i}{\pi}J^A_{-1}\log\frac{\rho}{2\Lambda} +
\left(\gamma - \frac{\rho^2}{2\gamma}\right)\log\frac{\gamma}{\Lambda}
+ O(\rho^4) + \ {non-log}\ terms,
\ee
where the first term at the r.h.s. is combined with the other
$J^A_{2m}\log\frac{\rho}{2\Lambda}$, and only the second one requires
further consideration.
Keeping all this in mind, we get for the

\be
{log-terms}\ in\ \frac{\Pi_i}{2g_{n+1}} = 
\frac{\hat S_i}{2}\log\frac{\rho_i}{2\Lambda} - 
\nn \\
-\frac{1}{2}\sum_{l\neq i}
\left[
\left(\rho_l^2 + \frac{\rho_l^4}{8}
\frac{\partial^2}{\partial\gamma_l^2}\right)
\varepsilon^{[il]}_{-1}(\gamma_{i\cdot}|\gamma_{l\cdot})
\left(\gamma_{il} - \frac{\rho_i^2}{2\gamma_{il}}\right)
\right]
\log\frac{\gamma_{il}}{\Lambda} + 
\nn \\
+ \frac{1}{4}\sum_{\stackrel{l_1<l_2}{l_1,l_2\neq i}}
\frac{\rho_{l_1}^2\rho_{l_2}^2}{\gamma_{l_1l_2}}
\left(
\varepsilon^{[il_1l_2]}_{-1}(\gamma_{i\cdot}|\gamma_{l_1\cdot})
\gamma_{il_1} \log\frac{\gamma_{il_1}}{\Lambda} -
\varepsilon^{[il_1l_2]}_{-1}(\gamma_{i\cdot}|\gamma_{l_2\cdot})
\gamma_{il_2} \log\frac{\gamma_{il_2}}{\Lambda}
\right) + O(\rho^6) =
\nn \\
\nn \\
=  \frac{\hat S_i}{2}\log\frac{\rho_i}{2\Lambda}
+ \frac{1}{2}\sum_{l\neq i}
\left[
\left(\rho_l^2 + \frac{\rho_l^4}{8}
\frac{\partial^2}{\partial\gamma_l^2}\right)
\tilde\Delta_l\left(1 - \frac{\rho_i^2}{2\gamma_{il}^2}\right)
\right]
\log\frac{\gamma_{il}}{\Lambda} -
\nn \\
-\frac{1}{4}\sum_{l\neq i}
\rho_l^2\tilde\Delta_l\log\frac{\gamma_{il}}{\Lambda}
\sum_{k\neq i,l}\frac{\rho_k^2}{\gamma_{lk}^2} + O(\rho^6) = 
\nn \\ \nn \\ =
\frac{\hat S_i}{2}\log\frac{\rho_i}{2\Lambda} +
\nn \\ 
+ \frac{1}{2}\sum_{l\neq i}\rho_l^2\tilde\Delta_l
\left(
1 + \frac{\rho_l^2}{4}\sum_{\stackrel{k_1<k_2}{k_1,k_2\neq l}}
\frac{1}{\gamma_{lk_1}\gamma_{lk_2}}  - 
\frac{1}{2}\frac{\rho_i^2}{\gamma_{li}^2}
 - \frac{1}{2}\sum_{k\neq i,l}\frac{\rho_k^2}{\gamma_{lk}^2}
 + O(\rho^6)
\right) \log\frac{\gamma_{il}}{\Lambda} =
\nn \\ \nn \\ \stackrel{(\ref{Srho})}{=}
\frac{\hat S_i}{4}\log\frac{\rho_i^2}{4}
- \frac{\hat S_i}{2}\log\Lambda + 
\frac{1}{2}\sum_{j\neq i}\hat S_j\log\frac{\gamma_{ij}}{\Lambda}
+ O(S^3) =
\nn \\ \nn \\
= -\frac{1}{2}(\sum_{j=1}^n \hat S_j) \log\Lambda 
+ \frac{\hat S_i}{4}\log\frac{\rho_i^2\Delta_i}{4}
- \frac{1}{4}\sum_{j\neq i}(\hat S_i-2\hat S_j)\log\gamma_{ij}
\label{logterms}
\ee
We see that the terms of order $\rho^4$ 
are nicely combined to correct $\rho_j^2\tilde\Delta_j$
into $\hat S_j$, according to (\ref{Srho}). The same
will, of course, happen in all higher orders of expansion
in powers of $\rho^2$, so that the answer at the r.h.s. 
of (\ref{logterms}) is in fact exact (but our calculation is
true only up to the $O(S^3)$ terms).
The first term at the r.h.s. of (\ref{logterms})
with $\log\rho_i^2$ is  combined with (\ref{fin0}) from the next
subsection to give an expression entirely in terms of $S$,
according to eq.(\ref{slog-s}).

\subsubsection{Polynomial terms}

The sources of the relevant (for calculation of the cubic
piece in the prepotential) terms in $\Pi_i/2g_{n+1}$ are:

-- the $\Lambda_i$-independent contributions (\ref{lambdaindep}) from 
terms with $Z_i$ (such contributions exist, because 
$Z_i \neq 2 \Lambda_i$) -- these  give

\be
-\frac{1}{4}e^{[i]}_{n-1}(\gamma_{i\cdot})\rho_i^2 = 
-\frac{\rho_i^2\tilde\Delta_i}{4}
\label{fin0}
\ee
in the order $O(\rho^2)$ and

\be
\frac{1}{32}e^{[i]}_{n-3}(\gamma_{i\cdot})\rho_i^4
+ \frac{1}{8}\rho_i^2\sum_{l\neq i}
\varepsilon^{[il]}_0(\gamma_{i\cdot}|\gamma_{l\cdot})\rho_l^2
= \nn \\ =
\frac{\rho_i^4\tilde\Delta_i}{32}
\sum_{\stackrel{j<k}{j,k\neq i}}\frac{1}{\gamma_{ij}\gamma_{ik}}
+ \frac{1}{8}\sum_{l\neq i}
\frac{\rho_i^2\rho_l^2(\tilde\Delta_i + \tilde\Delta_l)}{\gamma_{il}^2}
\label{fin1}
\ee
in the order $O(\rho^4)$;

-- the products of the non-derivative
term $\left(-\frac{1}{2}\rho_l^2\right)$ in 
operators $\hat D(\rho_l;\gamma_{il})$
and contribution $\left(-\frac{1}{4}\frac{\rho_i^2}{\gamma_{il}}\right)$ in
$J^B_{-1}(\rho_i,\gamma_{il})$, on which this operator
acts -- these give

\be
+ \frac{1}{8}\sum_{l\neq i} 
\varepsilon^{[il]}_{-1}(\gamma_{i\cdot}|\gamma_{l\cdot})
\frac{ \rho_i^2\rho_l^2}{\gamma_{il}}
= -\frac{1}{8}\sum_{l\neq i}\frac{\rho_i^2\rho_l^2\tilde\Delta_l}
{\gamma_{il}^2}; 
\label{fin2}
\ee

-- the results of application of the double-derivative term
$\left(-\frac{1}{16}\rho_l^4 \partial^2/\partial\gamma_{il}^2\right)$
to $\left(-\tilde\Delta_l\cdot\log\gamma_{il}\right)$
-- the $\rho$-independent contribution to 
$\varepsilon^{[il]}_{-1}(\gamma_{i\cdot}|\gamma_{l\cdot})
J^B_{-1}(\rho_i,\gamma_{il}) = \frac{\tilde\Delta_l}{\gamma_{li}}
J^B_{-1}(\rho_i,\gamma_{il})$  
-- these give

\be
+ \frac{1}{16}\sum_{l\neq i}\frac{\rho_l^4}{\gamma_{li}}
\left(2\frac{\partial\tilde\Delta_l}{\partial\gamma_l} 
- \frac{\tilde\Delta_l}{\gamma_{li}} 
\right)  = 
\frac{1}{8}\sum_{l\neq i}\frac{\rho_l^4\tilde\Delta_l}{\gamma_{li}}
\left(\sum_{k\neq i,l}\frac{1}{\gamma_{lk}} +
\frac{1}{2\gamma_{li}}\right)
\label{fin3}
\ee

All other terms are of different order in $\rho^2$:
the ones of lower order give rise to "universal" terms, forming
{\it perturbative} prepotential, the ones of higher order,
i.e. $O(\rho^6)$ contribute to higher non-perturbative corrections. 
Among the listed terms only (\ref{fin0}) is of order $O(\rho^2)$.
It gets combined with the logarithmic term 
$\frac{1}{4}\hat S_1\log\rho_1^2$, and the two together are
expressed through $\hat S_1$, with the help of eq.(\ref{slog-s}).

\subsection{Intermediate answer for the first non-perturbative correction}

The answer for the integral $\Pi_i/2g_{n+1}$ is represented as
a sum of two different kinds of terms: the "universal" ones (basically, the
cut-off dependent and {\it pertubative} contributions),

\be
W_{n+1}(\Lambda) - \frac{1}{2}\sum_i \hat S_i \log\Lambda +
\frac{1}{4}\hat S_i\log(\frac{1}{4}\hat S_i) - \frac{1}{4}\hat S_i - \nn \\
- W_{n+1}(\gamma_i) - \frac{1}{4}\sum_{j\neq i}
(\hat S_i - 2\hat S_j)\log \gamma_{ij}
\label{uniterms}
\ee
and the {\it non-perturbative} ones, of which the first non-trivial is
given by the sum of (\ref{fin1}), (\ref{fin2}) and (\ref{fin3}):

\be
\frac{\rho_i^4\tilde\Delta_i}{32}
\sum_{\stackrel{j<k}{j,k\neq i}}
\frac{1}{\gamma_{ij} \gamma_{ik}}
+ \frac{1}{8}\sum_{j\neq i}
\frac{\rho_i^2\rho_j^2(\tilde\Delta_i + \tilde\Delta_j)}{\gamma_{ij}^2}
-\frac{1}{8}\sum_{j\neq i}
\frac{\rho_i^2\rho_j^2 \tilde\Delta_j}{\gamma_{ij}^2} +
\nn \\
+ \frac{1}{8}\sum_{j\neq i}
\frac{\rho_j^4 \tilde\Delta_j}{\gamma_{ji}}\left(
\frac{1}{2\gamma_{ji}} + 
\sum_{k\neq i,j}\frac{1}{\gamma_{jk}}\right)
\label{f3terms}
\ee
(there is an obvious cancellation between the last term in the first line
and a piece of the second term).
Other contributions are of higher order $O(S^3)$ and can also be
straightforwardly evaluated.

To transform this formula into a final answer, one still needs
to get rid of remaining $\gamma$'s and express them through $\alpha$'s
and $S$'s. 

\section{Final formulas}

In quadratic order in $S$ all the $\gamma$'s in (\ref{f3terms})
can be just substituted by $\alpha$'s, since the differences
$\sigma_i = \gamma_i - \alpha_i = O(\rho^2) = O(S)$.
In (\ref{uniterms}) the only $\gamma$-dependencies  are in the second
line: in the argument of $W_{n+1}(\gamma_i)$
(note that the {\it coefficients} of $W(x) = \prod (x-\alpha_i)$ are already 
expressed through $\alpha$'s) and in $\log\gamma_{ij}$.
Both things are easy to control: according to
(\ref{sigmaS}) in the quadratic order in $S$,

\be
W_{n+1}(\gamma_i) = W_{n+1}(\alpha_i) +
\frac{1}{2}W_{n+1}^{\prime\prime}(\alpha_i)\sigma_i^2 + O(S^3) = 
\nn \\
= W_{n+1}(\alpha_i) + \frac{1}{8\Delta_i}\left(\sum_{j\neq i}
\frac{\hat S_i-\hat S_j}{\alpha_{ij}}\right)^2 + O(S^3)
\ee
and

\be
\log\gamma_{ij} = \log\alpha_{ij} + \frac{\sigma_{ij}}{\alpha_{ij}} 
+ O(S^2) =
\log\alpha_{ij} - 
\nn \\ 
-\frac{1}{2\alpha_{ij}}
\left[
\frac{\hat S_i-\hat S_j}{\alpha_{ij}}
\left(\frac{1}{\Delta_i}-\frac{1}{\Delta_j}\right) +
\sum_{k\neq i,j}
\left(
\frac{\hat S_i-\hat S_k}{\alpha_{ik}\Delta_i} -
\frac{\hat S_j-\hat S_k}{\alpha_{jk}\Delta_j}
\right)
\right]
\ee

The terms without $\sigma$'s, when substituted into (\ref{uniterms})
give rise to the "universal" terms, explicitly written in the expression 
(\ref{prepoexp}) for the prepotential. The terms with $\sigma$'s add to 
(\ref{f3terms}) to give:

\be
\frac{\hat S_i^2}{32\Delta_i}
\sum_{\stackrel{j<k}{j,k\neq i}}
\frac{1}{\alpha_{ij}\alpha_{ik}}
+ \frac{1}{8}\sum_{j\neq i}\frac{\hat S_i\hat S_j}{\alpha_{ij}^2\Delta_j}
+ \frac{1}{8}\sum_{j\neq i}\frac{\hat S_j^2}{\alpha_{ji}\Delta_j}
\left(+\frac{1}{2\alpha_{ji}} + \sum_{k\neq j,i}\frac{1}{\alpha_{jk}}\right)
- \nn \\ -
\frac{1}{8\Delta_i}\left(\sum_{j\neq i} \frac{\hat S_i-\hat S_j}{\alpha_{ij}}
\right)^2
+ \nn \\ +
\frac{1}{8}\sum_{j\neq i}\frac{\hat S_i-2\hat S_j}{\alpha_{ij}}
\left[
\frac{\hat S_i - \hat S_j}{\alpha_{ij}}\left(\frac{1}{\Delta_i} -
\frac{1}{\Delta_j}\right) +
\sum_{k\neq i,j}
\left(\frac{\hat S_i - \hat S_k}{\alpha_{ik}\Delta_i} -
\frac{\hat S_j - \hat S_k}{\alpha_{jk}\Delta_j}\right)
\right]
\ee

Finally, the coefficients in the formula for
$\Pi^{(3)}_i = (2\pi i)^{-1}\partial{\cal F}_3/\partial S_i$,

\be
\frac{1}{2g_{n+1}}\Pi^{(3)}_i =
v_i(\alpha)\hat S_i^2 + 
\sum_{j\neq i}v_{i;ij}(\alpha)\hat S_i \hat S_j + 
\sum_{j\neq i}v_{i;j}(\alpha)\hat S_j^2 +
\sum_{\stackrel{j<k}{j,k\neq i}}v_{i;jk}(\alpha)\hat S_j\hat S_k
\label{ans0}
\ee
are given by the following expressions:

\be
v_i = \frac{1}{8}
\left[
\frac{1}{4\Delta_i}
\sum_{\stackrel{j<k}{j,k\neq i}}
\frac{1}{\alpha_{ij}\alpha_{ik}}
- \frac{1}{\Delta_i}\left(\sum_{j\neq i}\frac{1}{\alpha_{ij}}\right)^2
+ \sum_{j\neq i}\frac{1}{\alpha_{ij}^2}
\left(\frac{1}{\Delta_i}-\frac{1}{\Delta_j}\right) + 
\right. \nn \\ \left. +
\frac{1}{\Delta_i}\sum_{j\neq i}\left(\sum_{k\neq i,j}
\frac{1}{\alpha_{ij}\alpha_{ik}}\right)
\right] = \nn \\ \nn \\ =
\frac{1}{8}
\left[
\frac{1}{4\Delta_i}
\sum_{\stackrel{j<k}{j,k\neq i}}
\frac{1}{\alpha_{ij}\alpha_{ik}}
- \sum_{j\neq i}\frac{1}{\alpha_{ij}^2\Delta_j}
\right]
\ee

\be
v_{i;ij} = 
\frac{1}{8}\left[
\frac{1}{\alpha_{ij}^2\Delta_j} + \frac{2}{\alpha_{ij}\Delta_i}
\left(\frac{1}{\alpha_{ij}} + \sum_{k\neq i,j}\frac{1}{\alpha_{ik}}\right)
- \frac{3}{\alpha_{ij}^2}\left(\frac{1}{\Delta_i} - \frac{1}{\Delta_j}\right)
- \right.\nn\\ \left. -
\frac{2}{\alpha_{ij}\Delta_i}\sum_{k\neq i,j}\frac{1}{\alpha_{ik}} -
\frac{1}{\alpha_{ij}\Delta_j}\sum_{k\neq i,j}\frac{1}{\alpha_{jk}}
- \frac{1}{\alpha_{ij}\Delta_i}\sum_{k\neq i,j}\frac{1}{\alpha_{ik}}
+ \sum_{k\neq i,j}\frac{1}{\alpha_{ik}\alpha_{kj}\Delta_k}
\right] = \nn \\ \nn \\ =
\frac{1}{8}\left[
-\frac{1}{\alpha_{ij}^2\Delta_i} + \frac{4}{\alpha_{ij}^2\Delta_j}
-\frac{1}{\alpha_{ij}\Delta_i}\sum_{k\neq i,j}\frac{1}{\alpha_{ik}}
-\frac{1}{\alpha_{ij}\Delta_j}\sum_{k\neq i,j}\frac{1}{\alpha_{jk}}
-\sum_{k\neq i,j} \frac{1}{\alpha_{ik}\alpha_{jk}\Delta_k}
\right]
= \nn \\ \nn \\ =
\frac{1}{8}\left[
-\frac{3}{\alpha_{ij}^2\Delta_i} + \frac{2}{\alpha_{ij}^2\Delta_j}
-\frac{2}{\alpha_{ij}\Delta_i}\sum_{k\neq i,j}\frac{1}{\alpha_{ik}}
\right] + \nn \\
+ \frac{1}{8}\left[
\frac{2}{\alpha_{ij}^2\Delta_i} + \frac{2}{\alpha_{ij}^2\Delta_j}
+\frac{1}{\alpha_{ij}\Delta_i}\sum_{k\neq i,j}\frac{1}{\alpha_{ik}}
-\frac{1}{\alpha_{ij}\Delta_j}\sum_{k\neq i,j}\frac{1}{\alpha_{jk}}
-\sum_{k\neq i,j} \frac{1}{\alpha_{ik}\alpha_{jk}\Delta_k}
\right]
= \nn \\ \nn \\ =
\frac{1}{8}\left[
-\frac{3}{\alpha_{ij}^2\Delta_i} + \frac{2}{\alpha_{ij}^2\Delta_j}
-\frac{2}{\alpha_{ij}\Delta_j}\sum_{k\neq i,j}\frac{1}{\alpha_{ik}}
\right]
\label{ans1}
\ee

\be
v_{i;j} = \frac{1}{8}\left[
\frac{1}{2\alpha_{ij}^2\Delta_j} - 
\frac{1}{\alpha_{ij}\Delta_j}\sum_{k\neq i,j}\frac{1}{\alpha_{jk}}
- \frac{1}{\alpha_{ij}^2\Delta_i} +
\right.\nn\\ \left. +
\frac{2}{\alpha_{ij}^2}\left(\frac{1}{\Delta_i} - \frac{1}{\Delta_j}\right)
+ \frac{2}{\alpha_{ij}\Delta_j}\sum_{k\neq i,j}\frac{1}{\alpha_{jk}}
\right] = \nn \\ \nn \\ =
\frac{1}{8}\left[
\frac{1}{\alpha_{ij}^2\Delta_i} - \frac{3}{2}\frac{1}{\alpha_{ij}^2\Delta_j}
+\frac{1}{\alpha_{ij}\Delta_j}\sum_{k\neq i,j}\frac{1}{\alpha_{jk}}
\right]
\label{ans2}
\ee
and

\be
v_{i;jk} = 
\frac{1}{8}\left[
-\frac{2}{\alpha_{ij}\alpha_{ik}\Delta_i} +
\frac{2}{\alpha_{ij}\alpha_{ik}\Delta_i} -
\frac{2}{\alpha_{ij}\alpha_{jk}\Delta_j} +
\frac{2}{\alpha_{ik}\alpha_{ij}\Delta_i} -
\frac{2}{\alpha_{ik}\alpha_{kj}\Delta_k}
\right] = \nn \\ \nn \\ =
\frac{1}{4}\left[\frac{1}{\alpha_{ij}\alpha_{ik}\Delta_i} +
\frac{1}{\alpha_{ji}\alpha_{jk}\Delta_j} +
\frac{1}{\alpha_{ki}\alpha_{kj}\Delta_k}\right]
\label{ans3}
\ee
Note, that integrability conditions for these relations are satisfied:

\be
v_{i;jk} = v_{j;ik} = v_{k;ij}
\ee
for all triples of different $i,j,k$
(of course, also $v_{i;jk} = v_{i;kj}$) and

\be
v_{i;ij} = 2v_{j;i}
\ee
for all $i\neq j$.
The last equation follows from somewhat non-trivial relation,

\be
\sum_{k\neq i,j} \frac{1}{\alpha_{ik}\alpha_{jk}\Delta_k} =
\frac{2}{\alpha_{ij}^2\Delta_i} + \frac{2}{\alpha_{ij}^2\Delta_j}
+\frac{1}{\alpha_{ij}}\sum_{k\neq i,j}
\left(\frac{1}{\alpha_{ik}\Delta_i}
- \frac{1}{\alpha_{jk}\Delta_j}\right)
\label{consist}
\ee
It is just a triviality for $n=2$, when only the first two terms survive
and $\Delta_1=-\Delta_2=\alpha_{12}$, while already for $n=3$ it already
requires some calculation:

\be
\frac{1}{\Delta_3^2} =
\frac{2}{\alpha_{12}^2\Delta_1} + \frac{2}{\alpha_{12}^2\Delta_2}
+ \frac{1}{\Delta_1^2}  + \frac{1}{\Delta_2^2} 
\ee
For generic $n$ it can be proved by showing that all the singularities
at $\alpha_{kl}=0$ cancel in the difference between the r.h.s. and l.h.s.
(but poles up to the third order are present and analysis of residues
is rather long).

From (\ref{ans0}-\ref{ans3}) it is easy to obtain the final 
expression (\ref{F3}) for the cubic term in the CIV-DV prepotential:

\be
\frac{u_i}{i\pi g_{n+1}} = \frac{(2\pi i)\cdot (2g_{n+1})}{(i\pi g_{n+1})^2}
\frac{v_i}{3}, \ \ i.e.\ \ 
u_i = \frac{4}{3}v_i, \nn \\
u_{i;j} = \frac{4}{2}v_{i;ij} = 4v_{j;i}, \nn \\
u_{ijk} = 4v_{i;jk}
\ee

\section{Conclusion}

Now eq.(\ref{F3}) awaits an independent check, also from perturbative
calculations for matrix models. A challenging problem is to further develop 
the machinery for evaluation of the higher-order terms in ${\cal F}_k$
in the prepotential. Simplicity of the answer (\ref{F3}) strongly suggests
that this should not be too difficult to do. A key point  may be to
understand the origins of identity (\ref{consist}) and its
generalizations  from the point of view of Seiberg-Witten theory
 or  even via a free fermion method.
It may provide a clue to the valuable tool to
the entire problem.

Coming back to (\ref{F3}), one of its immediate applications is to
experimental tests of the WDVV eqs \cite{MMM}, with \cite{IM4,IM5}
and {\it without} additional $T$-moduli. 

Obvious directions of the further analysis include:

-- introduction of {\it flat} $T$-moduli instead of $\alpha_{ij}$ \cite{IM4},

-- interpretation of above calculation in terms of free fermions, matrix
models and representation theory,

-- application to the $N=2$ SUSY prepotentials,

-- comparison to instanton calculus {\it a la} refs.\cite{Ne,Flu,gra},

-- generalization to numerous other models, of special interest being elliptic
ones and non-exactly-solvable (non-eigenvalue) matrix models.

\section{Acknowledgements}

A.M. acknowledges the support of JSPS and the hospitality
of the Osaka City University during his stay at Japan.
Our work is partly supported by the 
Grant-in-Aid for Scientific Research (14540284) from the
Ministry of Education, Science and Culture, Japan (H.I),
and by the Russian President's grant 00-15-99296, RFBR-01-02-17488,
INTAS 00-561 and by Volkswagen-Stiftung (A.M.).

\newpage

\appendix

\section{Relations between $\gamma_i$ and $\alpha_i$}

These relations follow from identification of (\ref{curve1})
and (\ref{curve2}),

\be
\prod_{L=1}^{2n} (x-\beta_L) \equiv
\prod_{i=1}^n(x-\gamma_i^-)(x-\gamma_i^+) =
\prod_{i=1}^n (x-\alpha_i)^2 + f_{n-1}(x),
\ee
where $\gamma_i^{\pm} = \gamma_i \pm \rho_i$.
The entire polynomial can be represented as

\be
\prod_{L=1}^{2n} (x-\beta_L) = 
\sum_{m=0}^{2n} (-)^me_m^{(2n)}\{\beta\}x^{2n-m},
\ee
and $f_{n-1}(x)$ does not contribute to the terms with $m\leq n$.
This means that 

\be
e_m^{(2n)}(\{\gamma^-\},\{\gamma^+\}) =
e_m^{(2n)}(\{\alpha\},\{\alpha\}), \ \ for\ 1\leq m\leq n.
\ee
At the same time, these symmetric polynomial of $2n$ variables can
be decomposed into bilinear combinations of those of $n$ variables,

\be
e_m^{(2n)}(\{\gamma^-\},\{\gamma^+\}) = \sum_{l=0}^m
e_l^{(n)}\{\gamma^-\}e_{m-l}^{(n)}\{\gamma^+\}
\ee
and

\be
e_m^{(2n)}(\{\alpha\},\{\alpha\}) = \sum_{l=0}^m
e_l^{(n)}\{\alpha\}e_{m-l}^{(n)}\{\alpha\},
\ee
and we obtain the "sum rules":

\be
\sum_{l=0}^m
e_l^{(n)}\{\alpha\}e_{m-l}^{(n)}\{\alpha\} =
\sum_{l=0}^m
e_l^{(n)}\{\gamma^-\}e_{m-l}^{(n)}\{\gamma^+\}
\ \ for\ 1\leq m\leq n.
\label{sumrules}
\ee
For given $n$ we have exactly $n$ such equations, which are
exactly what needed to express all $\gamma_i$ through 
$\alpha$'s and $\rho$'s. For higher $m>n$ the differences
between the l.h.s. and the r.h.s. of (\ref{sumrules}) provide
expressions for the coefficients of $f_{n-1}(x)$ through
$\alpha$'s and $\rho$'s.

The next step is to rewrite the sum rules (\ref{sumrules})
in terms of the more convenient basis of symmetric polynomial,
$p_m\{\alpha\} = \sum_{i=1}^n \alpha_i^m$:

\be
p_m\{\alpha\} = \frac{1}{2}\left(
p_m\{\gamma^-\} + p_m\{\gamma^+\}\right)
\ \ for\ 1\leq m\leq n.
\label{sumrules2}
\ee 
In particular, in terms of matrices\footnote{
Transition from (\ref{sumrules}) to (\ref{sumrules2}) is
provided by the usual trick, especially simple in 
matrix notation: for

$$
Z(\zeta; \check\alpha) \equiv 
\sum_{l=0}^n \zeta^l (-)^le_l^{(n)}(\{\alpha\}) = 
\prod_{l=0}^n (1-\zeta \alpha_i) = det_{n\times n}
(I - \zeta\check\alpha) =
$$
$$
= \exp\left(-\sum_{m=1}^\infty \frac{\zeta_m}{m}Tr\check\alpha^m\right)
= \exp\left(-\sum_{m=1}^\infty \frac{\zeta_m}{m} p_m\{\alpha\}\right)
$$
the sum rules (\ref{sumrules}) imply that

$$
Z(\zeta; \check\alpha)^2 = Z(\zeta; \check\gamma^-)
Z(\zeta; \check\gamma^+),
$$
for the first $n$ terms of expansion in powers of $\zeta$.
This, obviously, implies (\ref{sumrules2}).
}
$\check\alpha = diag(\alpha_1,\ldots,\alpha_n)$,
$\check\rho = diag(\rho_1,\ldots,\rho_n)$ and
$\check\gamma^\pm = diag(\gamma^\pm_1,\ldots,\gamma^\pm_n) =
\check\gamma \pm \check\rho$
(note, that $[\check\gamma, \check\rho] = 0$), 

\be
Tr\ \check\alpha =\ Tr\ \check\gamma, \nn \\
Tr\ \check\alpha^2 =\ Tr\ \check\gamma^2 + \ Tr\ \check\rho^2, \nn \\
Tr\ \check\alpha^3 =\ Tr\ \check\gamma^3 +  3Tr\ \check\gamma\check\rho^2, 
\nn \\
Tr\ \check\alpha^4 =\ Tr\ \check\gamma^4 +  6Tr\ \check\gamma^2\check\rho^2 
+ \ Tr\ \check\rho^4, \nn \\
Tr\ \check\alpha^5 =\ Tr\ \check\gamma^5 +  10Tr\ \check\gamma^3\check\rho^2 
+  5Tr\ \check\gamma\check\rho^4, \nn \\
\ldots
\label{sumrules3}
\ee

For {\bf $n=2$} the system of equations (\ref{sumrules3}) reduces to a single
quadratic equation, which can be solved exactly in all orders in
$\rho^2$:

\be
\gamma_{12} = \sqrt{\alpha_{12}^2-2(\rho_1^2 + \rho_2^2)}.
\ee
For {\bf $n=3$} the same can be done in terms of Cardano formulas for
solutions of cubic equations.

For generic $n$ one can get an
expansion in powers of $\rho$'s, by iteratively solving the system of 
equations for $\sigma_i \equiv \gamma_i - \alpha_i$,

\be
\ Tr\ \check\alpha^{m-1} \check\sigma =
-\frac{m-1}{2}\ Tr\ \check\alpha^{m-2}\check\rho^2 +
O(\rho^4)
\ee
($m=1,\ldots,n$).
 The determinant of the matrix  appearing on the left-hand side of these
equations is nothing but Van-der-Monde determinant
$\Delta = \prod_{i<j}^n \alpha_{ij}$.
Inverse of the Van-der-Monde matrix 
$\left(\alpha_i^{m-1}\right)$ is expressed through symmetric
polynomials:

\be
\sum_{m=1}^n (-)^{n-m-1}\frac{\Delta^{[i]}}{\Delta}e^{[i]}_{n-m}(\alpha)
\cdot\left(\alpha_j^{m-1}\right) =
\frac{\Delta^{[i]}}{\Delta} P_{[i]}(\alpha_j) = 
\delta_{ij} \frac{\Delta^{[i]}\Delta_i}{\Delta}
= \delta_{ij}
\ee
Here 

\be
P^{[i]}(x) \equiv \prod_{k\neq i}(x-\alpha_k)
= \frac{P(x)}{(x- \alpha_i)}
\ee
and

\be
\Delta^{[i]} = \prod_{\stackrel{j<k}{j,k\neq i}}\alpha_{jk} 
= \Delta \prod_{k\neq i}\frac{1}{\alpha_{ik}}, \ \ 
{\Delta_i} = \prod_{k\neq i}\alpha_{ik}
\ee
Therefore

\be
\sigma_i = -\frac{1}{2}\frac{\Delta^{[i]}}{\Delta}
\sum_{m=1}^n (-)^{n-m-1}e^{[i]}_{n-m}(\alpha)\cdot
(m-1)Tr \check \alpha^{m-2}\check\rho^2 = \nn \\ =
-\frac{1}{2\Delta_i} \sum_{j=1}^n P^{[i] \prime}(\alpha_j)\rho_j^2
= -\frac{1}{2}\prod_{k\neq i}\frac{1}{\alpha_{ik}}\left(
P^{[i] \prime}(\alpha_i)\rho_i^2 +
\sum_{j\neq i}P^{[i] \prime}(\alpha_j)\rho_j^2\right)
\ee
so that finally

\be
\sigma_i = -\frac{1}{2}\left(
\rho_i^2\sum_{k\neq i}\frac{1}{\alpha_{ik}} +
\sum_{j\neq i} \rho_j^2 \frac{\prod_{k\neq i,j}\alpha_{jk}}
{\prod_{k\neq i}\alpha_{ik}}\right)
+ O(\rho^4)
\label{sigmarho}
\ee
If we now use (\ref{zeta}) to express 

\be
\rho_{i}^2 = \hat S_i/\Delta_i + O(S^2),\ \ 
\Delta_i = \prod_{j\neq i}\alpha_{ij},
\ee
eq.(\ref{sigmarho}) acquires the simple final form:

\be
\sigma_i = -\frac{1}{2\Delta_i}\sum_{j\neq i}
\frac{\hat S_i- \hat S_j}{\alpha_{ij}} + O(S^2).
\label{sigmaS}
\ee

\section{Examples of $n=2$ and $n=3$}

For illustrative purposes we present here a more detailed
calculations for the simplest cases of $n=2$ and $n=3$.

\subsection{${\bf n=2:}$}

\be
-\frac{1}{2g_3}S_1 = A_0^{[1]}(\rho_1;\gamma_{12}) + 
\hat D(\rho_2;\gamma_{12})A_1^{[1]}(\rho_1;\gamma_{12}) = \nn \\ =
J^A_1(\rho_1) + \gamma_{12}J^A_0(\rho_1) +
\hat D(\rho_2;\gamma_{12}) J^A_{-1}(\rho_1,\gamma_{12}) = \nn \\ =
i\pi\left(c_1\rho_1^2\gamma_{12} + \sum_{k=1}^\infty c_k\rho_1^{2k}
\hat D(\rho_2;\gamma_{12})\frac{1}{\gamma_{12}^{2k-1}}\right) = \nn \\ =
-\frac{i\pi}{2}\left(
\rho_1^2\gamma_{12} - \sum_{k,l=0}^\infty
\frac{\rho_1^{2k+2}\rho_2^{2l+2}}{2^{2k+2l+1}}
\frac{(2k+2l)!}{k!(k+1)!l!(l+1)!}\frac{1}{\gamma_{12}^{2k+2l+1}}\right)
\label{2S1}
\ee
Similarly,

\be
-\frac{1}{2g_3}S_2 = A_0^{[2]}(\rho_2;\gamma_{21}) + 
\hat D(\rho_1;\gamma_{21})A_1^{[2]}(\rho_2;\gamma_{21}) = \nn \\ =
J^A_1(\rho_2) + \gamma_{21}J^A_0(\rho_2) +
\hat D(\rho_1;\gamma_{21}) J^A_{-1}(\rho_2,\gamma_{21}) = \nn \\ =
i\pi\left(c_1\rho_2^2\gamma_{21} + \sum_{l=1}^\infty c_l\rho_2^{2l}
\hat D(\rho_1;\gamma_{21})\frac{1}{\gamma_{21}^{2l-1}}\right) = \nn \\ =
-\frac{i\pi}{2}\left(
\rho_1^2\gamma_{21} - \sum_{k,l=0}^\infty
\frac{\rho_1^{2k+2}\rho_2^{2l+2}}{2^{2k+2l+1}}
\frac{(2k+2l)!}{k!(k+1)!l!(l+1)!}\frac{1}{\gamma_{21}^{2k+2l+1}}\right)
= \nn \\ =
\frac{1}{2g_3}S_1 - \frac{i\pi}{2}(\rho_1^2-\rho_2^2)\gamma_{12},
\label{2S2}
\ee
so that

\be
S_1+S_2 = i\pi g_3(\rho_1^2-\rho_2^2)\gamma_{12}
\ee

We do not write here an even more sophisticated general expressions for
$\Pi_1$ and $\Pi_2$: they can be easily obtained
in the form of {\it triple} series in powers of $\rho^2$, but after that one
still needs to express $\rho_1^2$ and $\rho_2^2$ through $S_1$ and $S_2$
from (\ref{2S1}) and (\ref{2S2}) and substitute into the series for
$\Pi_1$ and $\Pi_2$. However, the problem of inverting (\ref{2S1}) and 
(\ref{2S2}) is not resolved yet.

Instead, below we concentrate on the first terms of these expansions. 

\be
\hat S_1 = \frac{S_1}{i\pi g_3} = \rho_1^2\gamma_{12} -
\frac{1}{2}\frac{\rho_1^2\rho_2^2}{\gamma_{12}} + O(\rho^6), \nn \\
\hat S_2 = \frac{S_2}{i\pi g_3} = -\rho_2^2\gamma_{12} +
\frac{1}{2}\frac{\rho_1^2\rho_2^2}{\gamma_{12}} + O(\rho^6),
\ee

so that 
\be
\rho_1^2 = \frac{\hat S_1}{\gamma_{12}} - \frac{\hat S_1\hat S_2}{2\gamma_{12}^4}
+ O(\rho^6)
\nn \\
\rho_2^2 = -\frac{\hat S_2}{\gamma_{12}} - \frac{\hat S_1\hat S_2}{2\gamma_{12}^4}
+ O(\rho^6)
\ee

The $\Pi$ period is (remember that $D(\rho;\gamma_{ik})$
commutes with $\Lambda_i$!):

\be
\frac{\Pi}{2g_3} = J^B_1(\rho_1) + \gamma_{12}J^B_0(\rho_1) -
\frac{1}{2}\left(\rho_2^2 + \frac{\rho_2^4}{8}
\frac{\partial^2}{\partial \gamma_{12}^2}\right)
J^B_{-1}(\rho_1,\Lambda_1) = \nn \\
= \left(\frac{1}{24}Z_1^3 - \frac{1}{8}\rho_1^2Z_1\right)
+\left(
\frac{1}{2}\gamma_{12}\rho_1^2\log\frac{\rho_1}{2\Lambda} +
\frac{1}{8}\gamma_{12}Z_1^2\right)  \nn \\
- \frac{1}{2}\left(\rho_2^2 + \frac{\rho_2^4}{8}
\frac{\partial^2}{\partial \gamma_{12}^2}\right)
\left(\Lambda_1 + \gamma_{12}\log\frac{\gamma_{12}}{\Lambda}
+ \frac{\rho_1^2}{2\gamma_{12}}\log\frac{\rho_1}{2\gamma_{12}}
- \frac{\rho_1^2}{4\gamma_{12}}\right) + O(\rho^6) = 
\nn \\ \nn \\
= \frac{1}{3}(\Lambda^3-3\Lambda^2\gamma_1 + 3\Lambda\gamma_1^2 -
\gamma_1^3) + \frac{1}{2}\gamma_{12}(\Lambda^2-2\Lambda\gamma_1
+ \gamma_1^2) - \frac{1}{2}(\rho_1^2+\rho_2^2)(\Lambda-\gamma_1)
-\frac{1}{4}\gamma_{12}\rho_1^2 - 
\nn \\ 
- \frac{1}{2}\gamma_{12}(\rho_1^2-\rho_2^2)\log\Lambda 
+ \frac{1}{2}\gamma_{12}\rho_1^2\log\frac{\rho_1}{2} -
\frac{1}{2}\gamma_{12}\rho_2^2\log\gamma_{12}
- \frac{1}{4}\frac{\rho_1^2\rho_2^2}{\gamma_{12}}
\log\frac{\rho_1}{2\gamma_{12}} -
\nn \\
- \frac{\rho_2^4}{16\gamma_{12}} + \frac{\rho_1^2\rho_2^2}{8\gamma_{12}}
+ \nn \\ + O(\rho^6) =
\nn \\
\nn \\
= \frac{1}{3}(\Lambda^3-\gamma_1^3) 
- \frac{1}{2}(\gamma_1+\gamma_2)(\Lambda^2 - \gamma_1^2)
+\left(\gamma_1\gamma_2 - \frac{1}{2}(\rho_1^2+\rho_2^2)\right)(\Lambda - \gamma_1) -
\nn\\
- \frac{1}{2}\gamma_{12}(\rho_1^2-\rho_2^2)\log\Lambda 
+ \frac{1}{4}\left(\gamma_{12}\rho_1^2 -  
\frac{\rho_1^2\rho_2^2}{2\gamma_{12}}\right)\log\frac{\rho_1^2}{4} -
\frac{1}{2}\left(\gamma_{12}\rho_2^2 -  
\frac{\rho_1^2\rho_2^2}{2\gamma_{12}}\right)\log\gamma_{12}
- \nn \\
- \frac{1}{4}\gamma_{12}\left(\rho_1^2 -  
\frac{\rho_1^2\rho_2^2}{2\gamma_{12}}\right)
- \frac{\rho_2^4}{16\gamma_{12}}
+ \nn \\ + O(\rho^6) =
\nn \\
\nn \\
= W_3(\Lambda) - W_3(\gamma_1) - \frac{1}{2}(\hat S_1+\hat S_2)\log\Lambda
- \frac{1}{4}\hat S_1 +
\nn \\
+ \frac{1}{4}\hat S_1 \log\frac{\hat S_1}{4} 
+ \left(\frac{1}{2}\hat S_2 - \frac{1}{4}\hat S_1\right)\log\gamma_{12} -
\frac{\hat S_1\hat S_2}{8\gamma_{12}^3} - \frac{S_2^2}{16\gamma_{12}^3} 
+ O(S^3) 
\ee
At the last stage we
substituted the expression for $\rho_1^2$ through $\hat S_1$ and
$\hat S_2$ in the argument of the logarithm,

\be
\frac{1}{4}\hat S_1 \log\frac{\rho_1^2}{4} =
\frac{1}{4}\hat S_1 \left(\log\frac{\hat S_1}{4} - \log\gamma_{12} -
\frac{\hat S_2}{2\gamma_{12}^3}\right)
\ee
and moved the two newly emerging terms closer to the similar ones, coming from
other sources. Finally,

\be
\frac{\Pi}{2g_3} =
W_3(\Lambda)  - \frac{1}{2}(\hat S_1+\hat S_2)\log\Lambda
+\frac{1}{4}\hat S_1\log\frac{\hat S_1}{4} - \frac{1}{4}\hat S_1 -
\nn \\
- W_3(\gamma_1) - \frac{1}{4}(\hat S_1 - 2\hat S_2)\log\gamma_{12}
- \frac{\hat S_1\hat S_2}{8\gamma_{12}^3} - \frac{S_2^2}{16\gamma_{12}^3} 
+ O(S^3) 
\ee
Now, it remains to substitute

\be
W_3(\gamma_1) = W_3(\alpha_1) + \frac{1}{2}\alpha_{12}\sigma^2, \nn \\
\sigma = \gamma_1-\alpha_1 = - (\gamma_2-\alpha_2) = \frac{1}{2}
(\gamma_{12}-\alpha_{12}) = -\frac{\rho_1^2+\rho_2^2}{2\alpha_{12}} +
O(\rho^2) = \nn \\
= -\frac{\hat S_1-\hat S_2}{2\alpha_{12}^3} + O(S^2)
\ee
and

\be
\log\gamma_{12} = \log\alpha_{12} - \frac{\rho_1^2+\rho_2^2}{\alpha_{12}} +
O(\rho^2) = 
\log\alpha_{12} - \frac{\hat S_1-\hat S_2}{\alpha_{12}^3} + O(S^2)
\ee
to obtain:

\be
\frac{\Pi}{2g_3} =
W_3(\Lambda) - W_3(\alpha_1) - \frac{1}{2}(\hat S_1+\hat S_2)\log\Lambda
+\frac{1}{4}\hat S_1\log\frac{\hat S_1}{4} - \frac{1}{4}\hat S_1 +
\nn \\
+\frac{1}{8\alpha_{12}^3}\left(\hat S_1^2 - 5\hat S_1\hat S_2 + 
\frac{5}{2} \hat S_2^2\right)
\ee
(from
$2(S_1-S_2)(S_1-2S_2) - (S_1-S_2)^2 - S_1S_2 -\frac{1}{2}S_2^2$).

\subsection{${\bf n=3:}$}

The starting expression is

\be
\int_1 dS_{DV} = J_2(\rho_1;0) + (\gamma_{12}+\gamma_{13})J_1(\rho_1;0) + 
\gamma_{12}\gamma_{13}J_0(\rho_1;0)
+ \nn \\ +
\hat D(\rho_2,\gamma_{12})\left[J_0(\rho_1;0) + 
\gamma_{23}J_{-1}(\rho_1;\gamma_{12}|\Lambda_1)\right] +
\nn \\
+ \hat D(\rho_3,\gamma_{13})\left[J_0(\rho_1;0) - 
\gamma_{23}J_{-1}(\rho_1;\gamma_{13}|\Lambda_1)\right] 
+ \nn \\ +
\hat D(\rho_2,\gamma_{12})\hat D(\rho_3,\gamma_{13})
\frac{1}{\gamma_{23}}
\left[
J_{-1}(\rho_1;\gamma_{12}|\Lambda_1) - J_{-1}(\rho_1;\gamma_{13}|\Lambda_1)
\right]
+ O(\rho^6)
\ee

Expressions for $S$-integrals:

\be
-\frac{S_1}{2\pi ig_4} = -\frac{1}{8}\rho_1^4 -
\frac{1}{2}\gamma_{12}\gamma_{13}\rho_1^2 - \nn \\
-\frac{1}{2}\left(\rho_2^2 + 
\frac{\rho_2^4}{8}\frac{\partial^2}{\partial\gamma_{12}^2}\right)
\left(-\frac{1}{2}\rho_1^2 - \gamma_{23}\frac{\rho_1^2}{2\gamma_{12}}\right) -
\nn\\
-\frac{1}{2}\left(\rho_3^2 + 
\frac{\rho_3^4}{8}\frac{\partial^2}{\partial\gamma_{13}^2}\right)
\left(-\frac{1}{2}\rho_1^2 + \gamma_{23}\frac{\rho_1^2}{2\gamma_{13}}\right) +
\nn \\
+ \frac{1}{4}\rho_2^2\rho_3^2\frac{1}{\gamma_{23}}\left(
-\frac{\rho_1^2}{2\gamma_{12}} + \frac{\rho_1^2}{2\gamma_{13}}\right)
+ O(\rho^6)
\ee
In fact, not all of these terms actually contribute to the $S$-integrals
in the order $O(\rho^4)$ (however, their counterparts in the expressions
for $\Pi$-integrals {\it do} contribute). 
In result,
\be
\hat S_1 = \frac{S_1}{i\pi g_4} = 
\gamma_{12}\gamma_{13}\rho_1^2 + \frac{1}{4}\rho_1^4 -
\frac{\gamma_{13}}{2\gamma_{12}}\rho_1^2\rho_2^2 -
\frac{\gamma_{12}}{2\gamma_{13}}\rho_1^2\rho_3^2 + O(\rho^6)= \nn \\
= \gamma_{12}\gamma_{13}\rho_1^2
\left(1 +
\frac{\rho_1^2}{4\gamma_{12}\gamma_{13}} - \frac{\rho_2^2}{2\gamma_{12}^2} 
- \frac{\rho_3^2}{2\gamma_{13}^2}\right) + O(\rho^6)
\ee
Similarly, for $\hat S_2$ and $\hat S_3$.

Expressions for $\rho$'s through $\hat S_i = S_i/i\pi g_4$:

\be
\rho_1^2 = \frac{\hat S_1}{\gamma_{12}\gamma_{13}}
\left(1 - \frac{1}{4}\frac{\hat S_1}{\gamma_{12}^2\gamma_{13}^2}
+\frac{1}{2}\frac{\hat S_2}{\gamma_{21}^3\gamma_{23}} +
\frac{1}{2}\frac{\hat S_3}{\gamma_{31}^3\gamma_{32}}\right)
+ O(S^3); \nn\\
\rho_2^2 = \frac{\hat S_2}{\gamma_{21}\gamma_{23}}
\left(1 +\frac{1}{2}\frac{\hat S_1}{\gamma_{12}^3\gamma_{13}}
- \frac{1}{4}\frac{\hat S_2}{\gamma_{21}^2\gamma_{23}^2} +
\frac{1}{2}\frac{\hat S_3}{\gamma_{23}^3\gamma_{13}}\right)
+ O(S^3); \nn\\
\rho_3^2 = \frac{\hat S_3}{\gamma_{31}\gamma_{32}}
\left(1 + \frac{1}{2}\frac{\hat S_1}{\gamma_{13}^3\gamma_{12}}
+\frac{1}{2}\frac{\hat S_2}{\gamma_{23}^3\gamma_{21}} 
- \frac{1}{4}\frac{\hat S_3}{\gamma_{13}^2\gamma_{23}^2}\right)
+ O(S^3).
\ee

Now, the $\Pi$ integral:

\be
\frac{\Pi_1}{2g_4} = \frac{1}{64}Z_1^4(\rho_1) + 
\frac{1}{8}\rho_1^4\log\frac{\rho_1}{2\Lambda} + \nn \\
+(\gamma_{12}+\gamma_{13})\left(\frac{1}{24}Z_1^3(\rho_1) -
\frac{1}{8}Z_1(\rho_1)\right) + \nn \\
\gamma_{12}\gamma_{13}\left(\frac{1}{8}Z_1^2(\rho_1) +
\frac{1}{2}\rho_1^2\log\frac{\rho_1}{2\Lambda}\right) - \nn \\
-\frac{1}{2}\left(\rho_2^2 + 
\frac{\rho_2^4}{8}\frac{\partial^2}{\partial\gamma_{12}^2}\right)
\left[\left(\frac{\rho_1^2}{2}\log\frac{\rho_1}{2\Lambda} +
\frac{1}{8}Z_1^2(\rho_1)\right) + 
\right.\nn \\ \left.
+\gamma_{23}\left(\Lambda_1 + \gamma_{12}\log\frac{\gamma_{12}}{\Lambda}
+\frac{\rho_1^2}{2\gamma_{12}}\log\frac{\rho_1}{2\gamma_{12}}
- \frac{\rho_1^2}{4\gamma_{12}}\right)
\right] + \nn \\
-\frac{1}{2}\left(\rho_3^2 + 
\frac{\rho_3^4}{8}\frac{\partial^2}{\partial\gamma_{13}^2}\right)
\left[\left(\frac{\rho_1^2}{2}\log\frac{\rho_1}{2\Lambda} +
\frac{1}{8}Z_1^2(\rho_1)\right) -
\right.\nn \\ \left.
- \gamma_{23}\left(\Lambda_1 + \gamma_{13}\log\frac{\gamma_{13}}{\Lambda}
+\frac{\rho_1^2}{2\gamma_{13}}\log\frac{\rho_1}{2\gamma_{13}}
- \frac{\rho_1^2}{4\gamma_{13}}\right)
\right] + \nn \\
+\frac{\rho_2^2\rho_3^2}{4\gamma_{23}}
\left(
\gamma_{12}\log\frac{\gamma_{12}}{\Lambda} - 
\gamma_{13}\log\frac{\gamma_{13}}{\Lambda}
\right) + O(\rho^6)
\ee
where

\be
Z_1^4 = 16\Lambda_1^4 - 16\Lambda_1^2\rho_1^2 + 2\rho_1^4 + O(1/\Lambda), \nn \\
Z_1^3 = 8\Lambda_1^3 - 6\Lambda_1\rho_1 + O(1/\Lambda), \nn \\
Z_1^2 = 4\Lambda_1^2 - 2\rho_1^2 + O(1/\Lambda), \nn \\
Z_1 = 2\Lambda_1 + O(1/\Lambda).
\ee
Long transformation finally provides:\footnote{
Note that there are {\it no more} corrections to the cut-off-$\Lambda$-dependent 
terms, namely, no further contributions with higher powers of $\rho$'s. Still,
$\Lambda$ and $\gamma_i$ (isolated $\gamma_i$ not in the form of
 difference $\gamma_{ij}$) enter only as arguments of $W_4(x)$ with 
coefficients made
out of $\alpha$'s, not $\gamma$'s! This is made possible by {\it exact}
sum rules (\ref{sumrules2}), which do not get any higher-order corrections.
}

\be
\frac{\Pi_1}{2g_4} = W_4(\Lambda) - 
\frac{1}{2}(\hat S_1 + \hat S_2 + \hat S_3)\log\Lambda
+\frac{\hat S_1}{4}\log\frac{\hat S_1}{4} - \frac{\hat S_1}{4} - \nn \\
- W_4(\gamma_1) - \frac{1}{4}(\hat S_1-2\hat S_2)\log\gamma_{12} -
\frac{1}{4}(\hat S_1-2\hat S_3)\log\gamma_{13} + \nn \\
+ \frac{1}{32}\frac{\hat S_1^2}{\gamma_{12}^2\gamma_{13}^2}
+ \frac{\hat S_2^2}{16}\left(\frac{2}{\gamma_{12}^2\gamma_{23}^2}
- \frac{1}{\gamma_{12}^3\gamma_{23}}\right)
+ \frac{\hat S_3^2}{16}\left(\frac{2}{\gamma_{13}^2\gamma_{23}^2}
+ \frac{1}{\gamma_{13}^3\gamma_{23}}\right)
- \frac{1}{8}\frac{\hat S_1\hat S_2}{\gamma_{12}^3\gamma_{23}}
- \frac{1}{8}\frac{\hat S_1\hat S_3}{\gamma_{13}^3\gamma_{23}}
\nn \\
+ O(S^3,\Lambda^{-1})
\ee
In this expression the coefficients of $W'_4(x) = \prod_{i=1}^n(x-\alpha_i)$
are already expressed through $\alpha$. 
It remains to express $\gamma$'s through $\alpha$'s in the {\it argument}
of $W(\gamma_1)$ and in logarithmic terms -- all in the second line. 
In the given order in $S$, the $\gamma$'s in the third line can be
simply substituted by $\alpha$'s. This procedure literally repeats
the generic calculation in s.4 and we do not repeat it here.

\end{document}